\documentclass[11pt]{article}
\pdfoutput=1
\usepackage{fullpage}
\usepackage{cite}
\usepackage{graphicx}
\usepackage{amsmath}
\usepackage{amssymb}
\usepackage{subcaption}
\usepackage{color, colortbl}
\definecolor{Gray}{gray}{0.9}

\title{}
\date{}
\renewcommand{\vec}[1]{\mbox{\boldmath$ #1 $}}
\newcommand{\mat}[1]{\ensuremath{\bf{\textsf{#1}}}}
\def\beq{\begin{equation}}
\def\eeq{\end{equation}}
\allowdisplaybreaks
\begin{document}
\bibliographystyle{utphys}

\newcommand\n[1]{\textcolor{red}{(#1)}} 
\newcommand{\diff}{\mathop{}\!\mathrm{d}}
\newcommand{\lb}{\left}
\newcommand{\rb}{\right}
\newcommand{\f}{\frac}
\newcommand{\pd}{\partial}
\newcommand{\tr}{\text{tr}}
\newcommand{\fdiff}{\mathcal{D}}
\newcommand{\im}{\text{im}}
\let\caron\v
\renewcommand{\v}{\mathbf}
\newcommand{\T}{\tensor}
\newcommand{\R}{\mathbb{R}}
\newcommand{\C}{\mathbb{C}}
\newcommand{\Z}{\mathbb{Z}}
\newcommand{\msbar}{\ensuremath{\overline{\text{MS}}}}
\newcommand{\DIS}{\ensuremath{\text{DIS}}}
\newcommand{\abar}{\ensuremath{\bar{\alpha}_S}}
\newcommand{\bb}{\ensuremath{\bar{\beta}_0}}
\newcommand{\rc}{\ensuremath{r_{\text{cut}}}}
\newcommand{\Nd}{\ensuremath{N_{\text{d.o.f.}}}}
\def\dd{d\!\!{}^-\!}
\def\del{\delta\!\!\!{}^-\!}

\setlength{\parindent}{0pt}

\newcommand{\no}{\nonumber}
\renewcommand{\vec}[1]{\mbox{\boldmath$ #1 $}}
\renewcommand{\mat}[1]{\ensuremath{\bf{\textsf{#1}}}}

\titlepage
\begin{flushright}
QMUL-PH-20-24\\
\end{flushright}

\vspace*{0.5cm}

\begin{center}
{\bf \Large New heavenly double copies}

\vspace*{1cm} 
\textsc{
Erick Chac\'{o}n\footnote{e.c.chaconramirez@qmul.ac.uk}${}^a$,
Hugo Garc\'{\i}a-Compe\'an\footnote{compean@fis.cinvestav.mx}${}^b$,
Andr\'es Luna\footnote{luna@physics.ucla.edu}${}^c$, \\
Ricardo Monteiro\footnote{ricardo.monteiro@qmul.ac.uk}${}^a$,
  and Chris D. White\footnote{christopher.white@qmul.ac.uk}${}^a$
  } \\

\vspace*{0.5cm} $^a$ Centre for Research in String Theory, School of
Physics and Astronomy, \\
Queen Mary University of London, 327 Mile End
Road, London E1 4NS, UK\\

\vspace{0.5cm}
$^b$ {Departamento de F\'{\i}sica,}\\
Centro de Investigaci\'{o}n y de Estudios Avanzados
del Instituto Polit\'{e}cnico Nacional\\
{P.O. Box 14-740, CP. 07000, M\'exico D.F., M\'exico.}\\

\vspace*{0.5cm} $^c$ Mani L. Bhaumik Institute for Theoretical Physics, \\ UCLA Department of Physics and
Astronomy, Los Angeles, CA 90095, USA
\\

\end{center}

\vspace*{0.5cm}

\begin{abstract}
The double copy relates scattering amplitudes and classical solutions in
Yang-Mills theory, gravity, and related field theories.
Previous work has shown that this has an explicit realisation in self-dual YM theory, where the equation of motion can be written in a form that maps directly to Pleba\'nski's heavenly equation for self-dual gravity. 
The self-dual YM equation involves an
area-preserving diffeomorphism algebra, two copies of which appear in
the heavenly equation. In this paper, we show that this construction is a
special case of a wider family of heavenly-type examples, by (i) performing Moyal deformations, and (ii) replacing the
area-preserving diffeomorphisms with a less restricted algebra. As a result, we obtain a double-copy interpretation for hyper-Hermitian manifolds, extending the previously known hyper-K\"ahler case. We also introduce a double-Moyal deformation of the heavenly equation. The examples where the construction of Lax pairs is possible are manifestly consistent with Ward's conjecture, and suggest that the classical integrability of the gravity-type theory may be guaranteed in general by the integrability of at least one of two gauge-theory-type single copies.

\end{abstract}

\vspace*{0.5cm}

\section{Introduction}
\label{sec:intro}

A major focus of modern theoretical physics is the remarkable web of
connections between different field theories of interest. In this
paper, we explore one such connection, the so-called {\it double copy}
between gauge theory and gravity, which has been the basis of much
work over the past decade due to its applications. The double copy
originated in the study of perturbative scattering amplitudes, first
in string theory \cite{Kawai:1985xq} and later in quantum field theory
and gravity~\cite{Bern:2008qj,Bern:2010ue,Bern:2010yg}, where it has
been studied at various loop orders, both with and without supersymmetry. 
More recently, the notion of double-copy relations between theories
has been extended to the context of the classical limit of the
theories using a variety of approaches. These include relating exact, algebraically special
solutions to the equations of motion of the
theories~\cite{Monteiro:2014cda,Luna:2015paa,Luna:2016due,Carrillo-Gonzalez:2017iyj,Bahjat-Abbas:2017htu,Berman:2018hwd,Bah:2019sda,CarrilloGonzalez:2019gof,Banerjee:2019saj,Ilderton:2018lsf,Monteiro:2018xev,Luna:2018dpt,Lee:2018gxc,Cho:2019ype,Kim:2019jwm,Alfonsi:2020lub,Bahjat-Abbas:2020cyb,White:2016jzc,DeSmet:2017rve,Bahjat-Abbas:2018vgo,Elor:2020nqe,Gumus:2020hbb,Keeler:2020rcv,Arkani-Hamed:2019ymq,Huang:2019cja,Alawadhi:2019urr,Moynihan:2019bor,Alawadhi:2020jrv,Easson:2020esh,Casali:2020vuy,Cristofoli:2020hnk},
as well as applications to perturbative methods. Some examples of the
latter are constructing perturbative metrics directly \cite{Luna:2016hge}, solving the equations of motion in a world-line formalism
\cite{Goldberger:2016iau,Goldberger:2017frp,Goldberger:2017vcg,Goldberger:2017ogt,Shen:2018ebu,Carrillo-Gonzalez:2018pjk,Plefka:2018dpa,Plefka:2019hmz,Goldberger:2019xef,PV:2019uuv},
relating linearised solutions using a convolution
approach \cite{Anastasiou:2014qba,Borsten:2015pla,Anastasiou:2016csv,Cardoso:2016ngt,Borsten:2017jpt,Anastasiou:2017taf,Anastasiou:2018rdx,LopesCardoso:2018xes,Luna:2020adi,Borsten:2020xbt,Borsten:2020zgj},
as well as
applying double-copy ideas to the computation of quantities of
interest to gravitational wave astronomy, 
mainly in the post-Minkowskian
regime \cite{Luna:2017dtq,Kosower:2018adc,Maybee:2019jus,Bautista:2019evw,Bautista:2019tdr,Cheung:2018wkq,Bern:2019crd,Bern:2019nnu,Bern:2020buy,Kalin:2019rwq,Kalin:2020mvi,Almeida:2020mrg}. 
Closer to the idea of this paper, there is a large body of literature on webs of theories related by the double copy, e.g., \cite{Cachazo:2014xea,Johansson:2014zca,Johansson:2015oia,Chiodaroli:2015rdg,Chiodaroli:2015wal,Carrasco:2016ldy,Carrasco:2016ygv,Anastasiou:2017nsz,Azevedo:2018dgo}.
A comprehensive
review of these developments may be found in
refs.~\cite{Bern:2019prr,Borsten:2020bgv}.  \\

A crucial idea, which fostered the development of the double copy, is the {\it colour-kinematics duality} of Bern, Carrasco and Johansson (BCJ)~\cite{Bern:2008qj}. This `duality' remains a conjecture at loop level, but is well established for tree-level (i.e., classical) gauge theory. It states that the scattering amplitudes can be written in such a way that (apart from scalar-type propagators) the kinematic dependence has the same algebraic properties as the colour dependence, thus hinting that there is a kinematic analogue of the colour Lie algebra in gauge theory. In particular, there is a kinematic analogue for the Jacobi identity of the colour Lie algebra, which gives rise to a notion of {\it kinematic algebra}. Upon taking the double copy, the colour information is replaced by another copy of its kinematic analogue, so that two kinematic algebras appear in gravity
scattering amplitudes. In fact, starting from two distinct gauge theories (e.g., degree of supersymmetry or self-duality constraint), the two kinematic algebras in the gravity theory, each extracted from one of the `single copies', will be distinct. Despite some progress, the mathematical interpretation of a kinematic algebra remains elusive, partly because there is no understanding of the colour-kinematics duality at the level of the action or the equations of motion. This would rely on a formulation of gauge theory with a single cubic interaction vertex, whereas the usual Yang-Mills action has also a four-point vertex. This cubic vertex would have as colour factor the usual Lie algebra structure constant, $f^{abc}$, and as kinematic factor the putative structure constant of a kinematic algebra. \\

There is an exception, however, where the kinematic algebra has been fully understood. Reference~\cite{Monteiro:2011pc} circumvented the difficulties of the general  problem by restricting to the {\it self-dual sector} of pure Yang-Mills
theory. In a long-known formulation~\cite{Parkes:1992rz}, the action of self-dual YM involves a single adjoint-valued degree of freedom,
and is manifestly cubic. It is then possible
to completely elucidate the kinematic algebra alluded to above, which
turns out to correspond to certain area-preserving diffeomorphisms
(n.b.~in the mathematical literature, this group is often referred to as SDiff or SDiff${}_2$, and the algebra as sdiff or sdiff${}_2$). Furthermore, the equations of motion can be written in
terms of an interaction involving the product of two sets of structure
constants, for the colour and kinematic Lie algebras
respectively. Thus, the colour-kinematics duality becomes explicit, and the double copy to self-dual gravity straightforward. The equation of motion for self-dual gravity arises naturally in the form of Pleba\'nski's (second) heavenly equation \cite{Plebanski:1975wn}. \\

There have been many interesting results regarding kinematic algebras. They have been explored via algorithms for obtaining colour-kinematics-dual expressions for scattering amplitudes, e.g., in \cite{BjerrumBohr:2010zs,Mafra:2011kj,BjerrumBohr:2012mg,Fu:2012uy,Monteiro:2013rya,Mafra:2014oia,Naculich:2014rta,Mafra:2015vca,Bjerrum-Bohr:2016axv,Du:2017kpo,Edison:2020ehu}, and via a variety of algebraic and geometric approaches, e.g., in \cite{Cheung:2016prv,Fu:2016plh,Chen:2019ywi,Mizera:2019blq,Frost:2019fjn}. And yet, to date, it has not been possible to generalise the  very explicit construction for the self-dual sectors to the full Yang-Mills theory and gravity case. This motivates looking for examples of kinematic algebras that go beyond those of ref.~\cite{Monteiro:2011pc}. That is the main aim of our paper, and we will seek to generalise the results of ref.~\cite{Monteiro:2011pc} in two key ways. Firstly, we may form qualitatively different types of gauge and gravity theories by replacing the Lie and Poisson brackets appearing in the former and latter with their {\it Moyal deformations} \cite{Moyal:1949sk}. These arise in alternative formulations of quantum mechanics~\cite{Groenewold:1946kp,Baker:1958zza} as well as in the study of field theories in non-commutative spacetimes; see e.g. refs.~\cite{doi:10.1142/5287,li2002physics} for pedagogical reviews. The Moyal bracket constitutes the most general bracket one can write down for a Lie algebra of functions of two variables~\cite{Fletcher:1990ib}, here corresponding to the two-dimensional space on which the diffeomorphisms act. Secondly, we will consider replacing the sdiff algebra discussed above with the set of arbitrary -- rather than area-preserving -- two-dimensional diffeomorphisms (diff). Both of the above generalisations lead to new instances of the double copy, in which the colour-kinematics duality is manifest. These results are summarised in the table of theories of section~\ref{sec:summary}.\\

A second motivation for our paper is to explore the classical integrability (or otherwise) of the gauge and gravity theories related by the double copy, an aspect that was not considered in ref.~\cite{Monteiro:2011pc}. As is well-known, both self-dual Yang-Mills theory and self-dual gravity are integrable, and thus admit an infinite number of conserved charges. This integrability can be expressed in terms of a {\it Lax pair}, and we will find a hitherto unexplored double copy interpretation of Lax pairs in different theories. This suggests that integrability of a gravity theory is ``inherited'' from its corresponding gauge theories via the double copy. Also, in considering the more general theories outlined above, we will find that integrability in the gravity theory can be obtained even if only one of the gauge theories in the double copy is integrable.
\\

The structure of our paper is as follows. In section~\ref{sec:review}, we review the self-dual double copy, and relate this to ideas regarding integrability. We consider Moyal deformations in section~\ref{sec:Moyal}. In section~\ref{sec:diff}, we look at generalising the algebra of diffeomorphisms in the gauge theory, and construct a corresponding gravity theory.  We summarise the web of theories studied in this paper in section~\ref{sec:summary}, and the general form of the different Lax pairs and the infinite tower of linearised symmetries is examined as well. Finally, we discuss our results and conclude in section~\ref{sec:conclude}.

\section{The self-dual double copy}
\label{sec:review}

\subsection{Review}

In this section, we review salient details of the
double copy in the self-dual sector described in ref.~\cite{Monteiro:2011pc}. This will set up useful
notation for what follows, whilst also allowing us to draw attention
to aspects of integrability, which were not examined in \cite{Monteiro:2011pc}. Our starting point is to consider pure Yang-Mills
theory. The (vacuum) equation of motion is given by
\begin{equation}
D^\mu F_{\mu\nu}=0,
\label{YM} 
\end{equation}
where
\begin{equation}
F_{\mu\nu}=\partial_\mu A_\nu-\partial_\nu A_\mu-ig[A_\mu,A_\nu]
\label{Fdef}
\end{equation}
is the field strength tensor in terms of the gauge field $A_\mu\equiv
A^a_\mu\,{T}^a$, and ${T}^a$ is a generator of the Lie algebra ${\mathbf g}$ of the gauge group $G$. Also, $D_\mu=\partial_\mu-igA_\mu$
is the covariant derivative, where the gauge field in the second term
acts in the adjoint representation. The equations of motion take a
particularly simple form if one considers self-dual solutions, for
which the field strength tensor satisfies
\begin{equation}
F_{\mu\nu}=\frac{i}{2}\epsilon_{\mu\nu\rho\sigma} F^{\rho\sigma}.
\label{FSD}
\end{equation}
One may introduce coordinates
\begin{equation}
u=t-z,\quad v=t+z,\quad w=x+iy,\quad \bar{w}=x-iy,
\label{coords}
\end{equation}
such that the Minkowski line element becomes
\begin{equation}
ds^2=dudv-dwd \bar{w},
\label{ds2}
\end{equation}
and the self-duality conditions can be written as
\begin{gather}
F_{uw}=0,\label{one}\\
F_{v\bar{w}}=0,\label{two}\\
F_{uv}-F_{w\bar{w}}=0.\label{three}
\end{gather}
From eq.~(\ref{one}) and a light-cone gauge choice, we may take
\begin{equation}
A_{u}=A_{w}=0.
\label{Auw} 
\end{equation}
It follows from eq.~(\ref{three}) that there must exist an
adjoint-valued scalar function $\Phi\equiv \Phi^a {T}^a$ such that
\begin{equation}
A_{v}=-\partial_{w}\Phi,\quad 
A_{\bar{w}}=-\partial_{u}\Phi.
\label{Avwbar}
\end{equation}
Physically, $\Phi$ represents the single polarisation state that
remains in the gauge field upon projecting to the self-dual sector,
and eq.~(\ref{two}) implies that it satisfies the following equation
of motion:
\begin{equation}
\partial^2\Phi+ig[\partial_w\Phi,\partial_u\Phi]=0, 
\label{YMEOM}
\end{equation}
where
$\partial^2\equiv\partial_u\partial_v-\partial_w\partial_{\bar{w}}$.
We follow Ref. \cite{Monteiro:2011pc}, to rewrite this as an integral equation in momentum space. Fourier transforming
eq.~(\ref{YMEOM}) and rearranging yields
\begin{align}
\Phi(k)&=-ig\int d^D x e^{-ix\cdot (p_1+p_2-k)}
\int\dd p_1\int \dd p_2\frac{1}{k^2}
\left(p_{1w}p_{2u}-p_{1u}p_{2w}\right)\Phi(p_1)\Phi(p_2)\notag\\
&=-ig\int \dd p_1\dd p_2\frac{\del(p_1+p_2-k)}{k^2}
X(p_1,p_2)\Phi(p_1)\Phi(p_2),
\label{SDYMFourier}
\end{align}
where we have employed the shorthand notation 
\begin{equation}
\dd p\equiv \frac{d^D p}{(2\pi)^D}, \qquad 
\del(p)\equiv (2\pi)^D\delta^{(D)}(p),
\label{dddef}
\end{equation}
and, in the second line, have defined the kinematic structure constant
\begin{equation}
X(p_1,p_2)=p_{1w}p_{2u}-p_{1u}p_{2w}.
\label{Fp1p2def}
\end{equation}
Antisymmetry of the latter
under $p_1\leftrightarrow p_2$ means that one may replace the product
of scalar fields with a commutator, to finally write
\begin{equation}
\Phi^a(k)=\frac{g}{2}\int\dd{p}_1\,\dd{p}_2\,\frac{\del(p_1+p_2-k)}{k^2}
\,X(p_1,p_2)\,f^{a b c}\,\Phi^{b}(p_1)\Phi^{c}(p_2).
\label{YMsol}
\end{equation}
The $f^{abc}$ are the structure constants of the Lie algebra
${\mathbf g}$, $[T^{a},T^{b}]=if^{abc}T^{c}$.
The kinematic objects in eq. (\ref{Fp1p2def}) are also the structure constants of a Lie algebra, the Jacobi identity being
\begin{equation}
X(p_1,p_2)\,X(p_1+p_2,p_3)+
X(p_2,p_3)\,X(p_2+p_3,p_1)+
X(p_3,p_1)\,X(p_3+p_1,p_2)=0.
\label{XJacobi}
\end{equation}
The kinematic Lie algebra is infinitely dimensional, with generators of the form
\begin{equation}
L_k=e^{-ik\cdot x}(-k_w\partial_u+k_u\partial_w),
\label{Lkdef}
\end{equation}
satisfying
\begin{equation}
[L_{p_1},L_{p_2}]=iX(p_1,p_2) L_{p_1+p_2}.
\label{Lalgebra}
\end{equation}
It is the Lie algebra sdiff of area-preserving diffeomorphisms in
the $(w,u)$ plane. We can also interpret it as a Poisson algebra, with
\begin{equation}
\{e^{-ip_{1}x},e^{-ip_{2}x}\}=-X(p_{1},p_{2})\,e^{-i(p_{1}+p_{2})x},
\label{expbracket}
\end{equation}
where the Poisson bracket is 
\begin{equation}
\{A,B\}=\partial_w A\, \partial_u B -\partial_u A \,\partial_w B. 
\label{Poissonbracket}
\end{equation}
We see that the self-dual YM equation
(\ref{YMsol}) is precisely that expected from a Lagrangian
involving a cubic interaction only, whose Feynman rule involves the
product of structure constants for two Lie algebras, corresponding respectively to
colour, $f^{abc}$, and kinematics, $X(p,q) $. This makes
manifest the BCJ duality between colour and kinematics of
ref.~\cite{Bern:2008qj}, in a sector where the kinematic algebra is a straightforward Lie algebra.\\

\begin{figure}
\begin{center}
\scalebox{0.8}{\includegraphics{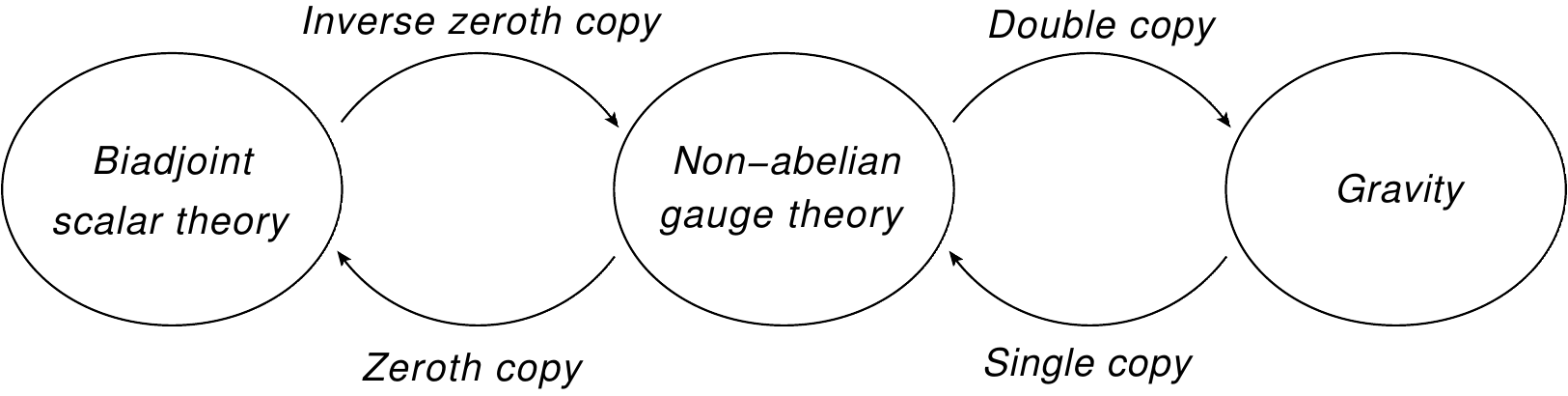}}
\caption{Various field theories and the relationships between
  them. Figure taken from ref.~\cite{Bahjat-Abbas:2020cyb}.}
\label{fig:theories}
\end{center}
\end{figure}

Given eq.~(\ref{YMsol}) and the interpretation given above, one may replace the colour structure constants with a second set of kinematic structure constants, i.e.
\begin{eqnarray}
f^{abc}\rightarrow X(p_1,p_2),
\label{DoubleCopy}
\end{eqnarray}
obtaining the momentum-space equation of motion
\begin{equation}
\phi(k)=\frac{\kappa}{2}\int\dd p_1\dd p_2\,\frac{\del(p_1+p_2-k)}{k^2}
\,X(p_1,p_2)^2\,\phi(p_1)\phi(p_2),
\label{Gravsol} 
\end{equation}
for a scalar field $\phi$. We now identify the coupling constant as the gravitational coupling $\kappa=\sqrt{32\pi G_N}$, in terms of Newton's constant $G_N$. In coordinate space, eq.~(\ref{Gravsol}) is simply the {\it Pleba\'nski equation } for self-dual gravity, also known as the  (second) {\it heavenly equation}:
\begin{equation}
\partial^2\phi+\kappa\{\partial_w\phi,\partial_u\phi\}=0.
\label{Plebanski}
\end{equation}
Thus, the double copy is particularly clear, and amounts to simply
replacing the colour algebra with its kinematic counterpart. One can
also go the other way in eq. (\ref{DoubleCopy}), replacing the area-preserving diffeomorphisms Lie algebra with a second `colour' Lie algebra $\tilde {\mathbf g}$
associated to a Lie group $\tilde{G}$. In this case, one
obtains solutions of the so-called {\it biadjoint scalar field
  theory}, whose equation of motion is
\begin{equation}
\partial^2\Phi^{aa'}+yf^{abc}\tilde{f}^{a'b'c'}\Phi^{bb'}\Phi^{cc'}=0,
\label{Phiaa'}
\end{equation}
or, in momentum space,
\begin{equation}
\Phi^{aa'}(k)=y\,\int\dd p_1\dd p_2\,\frac{\del(p_1+p_2-k)}{k^2}
\,f^{abc}\,\tilde{f}^{a'b'c'}\,\Phi^{bb'}(p_1)\Phi^{cc'}(p_2).
\label{BAdjsol} 
\end{equation}
Here, $y$ is a coupling constant, and $f^{abc}$, $\tilde{f}^{a'b'c'}$ are structure constants for the two Lie groups. There is thus a hierarchy
of theories as shown in figure~\ref{fig:theories}, which in the
self-dual sector involves scalar field theories with a cubic interaction. In
moving to the right in the figure, one removes an adjoint index from
the field, and replaces a set of colour structure constants in the
field equation with a set of kinematic structure constants. The biadjoint scalar theory plays a crucial role in the formulas expressing the double copy of scattering amplitudes. \\
From now on, we set the different coupling constants $(y,g,\kappa)$ to one.
\\

This `heavenly' example of the double copy is a different viewpoint to the well-known story in the integrability literature where self-dual
gravity is recovered as a symmetry reduction of
self-dual Yang-Mills theory, considering the gauge group as the
area-preserving diffeomorphism group on the two-surface $\Sigma$,
SDiff($\Sigma$)
\cite{Plebanski:1995gk,plebanski1994selfdual,park19922d}. This fact
has been used to build hyper-K\"ahler metrics from solutions of
two-dimensional reduced SDYM equations
\cite{Chacon:2018nff,Dunajski_1998,plebanski1996principal}. The double copy motivates the consideration of the biadjoint scalar theory mentioned above. Notice that self-dual Yang-Mills theory is a `symmetry reduction' of the biadjoint scalar theory in the same way that self-dual
gravity is a symmetry reduction of
self-dual Yang-Mills theory.  \\

Let us also comment on an alternative description of the self-dual theories. Whereas the second heavenly equation is eq. (\ref{Plebanski}), the first heavenly equation reads
\begin{equation}
\{\partial_v\Omega_{},\partial_{\bar{w}}\Omega_{}\}=1,
\label{FirstHeavenly}
\end{equation}
where $\Omega$ is the K\"ahler potential, and we use the $(w,u)$ Poisson bracket previously defined.
The two heavenly equations are equivalent and arise due to distinct gauge choices \cite{Plebanski:1975wn}.
A somewhat analogous formulation is possible in self-dual Yang-Mills, where instead of \eqref{Avwbar},  the relevant equation is
\begin{equation}
[\partial_{w}\Xi,\partial_{u}\Xi]=0,
\label{SigmaModel:equation}
\end{equation}
where $\Xi$ is Lie algebra-valued. This equation arises from a two-dimensional sigma model with pure SU(N) Wess-Zumino term, and is related to eq.~(\ref{FirstHeavenly}) in the large-N limit \cite{PARK1990287}, for reasons similar to those discussed in section~\ref{sec:DSDGvsSDG}. 
%
While equations (\ref{FirstHeavenly}) and \eqref{SigmaModel:equation} are closely related, the double copy structure is not at all clear, as in the case of (\ref{Plebanski}) and \eqref{YMEOM}, which are based on a cubic interaction vertex, the preferred setting for making the double copy manifest. Still, it would be interesting to explore these and other alternative formulations, in the search for further insights into the double copy.\\


\subsection{Lax pairs}
\label{sec:Lax}

There is an additional structure in the equations of motion \eqref{YMEOM} and (\ref{Plebanski}) for the self-dual theories, that did not receive attention in ref.~\cite{Monteiro:2011pc}, but which will be useful to review for what follows. Namely, both self-dual Yang-Mills and gravity are known to be integrable theories, admitting an infinite number of conserved charges. The existence of a Lax pair indicates that a given theory may be integrable, and we will be interested in tracing how integrability is inherited by different theories in our double copy examples. For self-dual YM,
this Lax pair consists of two operators obtained from the covariant derivative,
\begin{equation}
L=D_u-\lambda D_{\bar{w}},\quad M=D_w-\lambda D_v,
\label{LaxYM}
\end{equation}
where $\lambda\in{\mathbb CP}^1$ is an arbitrary {\it spectral
  parameter}. It is then straightforward to show that the self-dual
Yang-Mills equations of eqs.~(\ref{one}--\ref{three}) follow from the
compatibility condition
\begin{equation}
[L,M]=0.
\label{Laxbracket}
\end{equation}
Note that for the explicit gauge choice of eqs.~(\ref{Auw},
\ref{Avwbar}), the above Lax pair can also be written as
\begin{equation}
L=\partial_{u}-\lambda(\partial_{\bar{w}}+i\partial_{u}\Phi),\qquad\qquad M=\partial_{w}-\lambda(\partial_{v}+i\partial_{w}\Phi).
\label{LaxGaugeRedSD}
\end{equation}

The gravitational case can be expressed in a similar way. We can use known results to classify the types
of manifolds that arise, based on
the relevant Lax pair, as in \cite{Mason:1989ye,Dunajski:1998nj}. More specifically, let us take ${\mathbf
  V}=\{U,V,W,\bar{W}\}$ to be four independent holomorphic vector fields
on a four-dimensional complex manifold ${\cal M}$, such that
\begin{equation}
L=U-\lambda\bar{W},\quad M=W-\lambda V,\quad \lambda\in{\mathbb CP}^1,
\label{Lax:Gravity}
\end{equation}
are operators satisfying $\mathcal{L}_{L}\upsilon=\mathcal{L}_{M}\upsilon=0$ and eq. (\ref{Laxbracket}), for all $\lambda$, where $\upsilon$ is a non-zero holomorphic four-form and $\mathcal{L}$
denotes the Lie derivative. Then ${\bf V}$
forms a null tetrad for a so-called {\it hyper-K\"{a}hler} metric on
${\cal M}$.
To make contact with the Pleba\'nski equation, we can choose coordinates such that
\begin{align}
\nonumber U&=\partial_u, & V&=\partial_v+{\phi_{ww}}\partial_u-{ \phi_{uw}}\partial_w,\\
W&=\partial_w, & \bar{W}&=\partial_{\bar{w}}+{
\phi_{uw}}\partial_u-{\phi_{uu}}\partial_w,
\label{PleVec2}
\end{align}
which corresponds to
\begin{equation}
L=\partial_u-\lambda\left(\partial_{\bar{w}}+{\phi_{uw}}\partial_u-{\phi_{uu}}\partial_w \right), \quad M=\partial_w-\lambda\left(\partial_v+{\phi_{ww}}\partial_u-{\phi_{uw}}\partial_w \right),
\label{Laxgrav}
\end{equation}
where we use the notation $\phi_{\mu\nu}\equiv \partial_{\mu}\partial_{\nu}\phi$.
This Lax pair satisfies $[L,M]=0$ if $\phi$ satisfies the Pleba\'nski equation~(\ref{Plebanski}). Furthermore, by considering the holomorphic four-form
\begin{eqnarray}
\upsilon=du\wedge dv\wedge dw\wedge d\bar{w},
\end{eqnarray}
the Lax pair satisfies also the conditions $\mathcal{L}_{L}\upsilon=\mathcal{L}_{M}\upsilon=0$. Then, the vector fields in eq.~(\ref{PleVec2}) form a null tetrad for a hyper-K\"{a}hler metric.\\


The Lax pairs \eqref{LaxGaugeRedSD} and \eqref{Laxgrav} are naturally related, as expected by the double copy. Let us introduce the Hamiltonian vector field in the $(u,w)$ plane, $\Phi_{f}$, defined by its action on functions\footnote{This Hamiltonian vector field is defined in the $(u,w)$ plane. However, the Hamiltonian functions depend on the coordinates $(u,v,w,\bar{w})$.}, 
\begin{equation}
\Phi_{f}(g)\equiv\{f,g\},
\label{Poisson}
\end{equation}
so that
\begin{equation}
[\Phi_{f},\Phi_{g}]=\Phi_{\{f,g\}}.
\label{Poisson2}
\end{equation}
Then, the Lax pair in self-dual gravity, in eq.~\eqref{Laxgrav}, takes the form
\begin{equation}
L=\partial_{u}-\lambda(\partial_{\bar{w}}+\partial_{u}\Phi_{\phi}),\qquad 
M=\partial_{w}-\lambda(\partial_{v}+\partial_{w}\Phi_{\phi}),
\label{Laxgrav2}
\end{equation}
where the derivatives act only on the components of the Hamiltonian vector fields, that is, we define $\partial_u\Phi_{\theta_{n}}\equiv [\partial_u,\Phi_{\theta_{n}}]$ and $\partial_{w}\Phi_{\phi} \equiv[\partial_{w},\Phi_{\phi}]$. 
This Lax pair mirrors the one in self-dual YM, eq.~\eqref{LaxGaugeRedSD}. In the self-dual gravity case, we can also write the Lax pair as
\begin{equation}
L=\partial_{u}-\lambda(\partial_{\bar{w}}+\Phi_{\partial_{u}\phi}),\qquad 
M=\partial_{w}-\lambda(\partial_{v}+\Phi_{\partial_{w}\phi}).
\label{Laxgrav3}
\end{equation}


\subsection{Infinite tower of conserved currents}
\label{sec:Tower}
Let us return to the question of integrability. With the Lax pairs in hand, it is possible to construct recursively (formally at least) an infinite number of symmetries, associated to the infinite-dimensional vector space of deformations of solutions to the equations of motion. This brief review is roughly based on the discussions in \cite{Dunajski:2010zz,Cangemi:1996rx}. Let us take the self-dual YM example, with Lax pair given in \eqref{LaxGaugeRedSD}. Notice that
\begin{eqnarray}
[\partial_v+i\partial_w\Phi,L\Psi]-[\partial_{\bar{w}}+i\partial_u\Phi,M\Psi] =
\partial^2 \Psi +i [\partial_w\Phi,\partial_u\Psi]+i [\partial_w\Psi,\partial_u\Phi]\,,
\end{eqnarray}
where we have used the Jacobi identity and the equation of motion \eqref{YMEOM}.
Hence, we obtain a solution $\Psi$ to the linearised equation of motion, $\Phi \mapsto \Phi+\Psi$, by solving the system
\begin{eqnarray}
L\Psi=0,\qquad M\Psi=0\,.
\label{eq:LM0}
\end{eqnarray}
The compatibility condition of this overdetermined system is $[L,M]=0$. Now, let us consider a solution to eq. \eqref{eq:LM0} of the form
\begin{eqnarray}
\Psi(x,\lambda)=\sum_{n=0}^\infty \Psi_n(x)\lambda^n\,,
\end{eqnarray}
where $\lambda$ is the spectral parameter. Since
\begin{align}
& L\Psi= \sum_{n=0}^\infty \left(\partial_u\Psi_n-\left(\partial_{\bar{w}}\Psi_{n-1}+i[\partial_u\Phi,\Psi_{n-1}]\right)\right)\lambda^n,
\nonumber \\
& M\Psi= \sum_{n=0}^\infty \left(\partial_w\Psi_n-\left(\partial_{v}\Psi_{n-1}+i[\partial_w\Phi,\Psi_{n-1}]\right)\right)\lambda^n,
\end{align}
with $\Psi_{-1}\equiv0$,
the system \eqref{eq:LM0} is solved by an infinite tower of linearised solutions $\{\Psi_n\}$ constructed recursively via
\begin{align}
\partial_u\Psi_n=\partial_{\bar{w}}\Psi_{n-1}+i[\partial_u\Phi,\Psi_{n-1}],
\qquad 
\partial_w\Psi_n=\partial_{v}\Psi_{n-1}+i[\partial_w\Phi,\Psi_{n-1}].
\label{HierarchySDYM}
\end{align}
For instance, we can take $\Psi_0=\Psi_{0}^{a}T^a$, where $\Psi_{0}^{a}$ are constants, so that $\Psi_1=i[\Phi,\Psi_{0}^{a}T^a]$, etc. We can also express this hierarchy of infinitesimal symmetries in terms of a hierarchy of conserved currents, by considering
\begin{eqnarray}
J(x,\lambda)=(\partial_w\Psi) \partial_u - (\partial_u\Psi) \partial_w ,
\qquad \text{such that} \quad [\partial_\mu ,J^\mu]=0.
\end{eqnarray}
The infinite tower of currents is given by the $\lambda$-expansion of $J^\mu$. Equivalently, we can state that, if \eqref{HierarchySDYM} represents commuting flows of $\Psi_n$ along $u$ and $w$ (this is the compatibility condition $\partial_{wu}\Psi_n=\partial_{uw}\Psi_n$), then $\Psi_n$ is a linearised symmetry due to
\begin{eqnarray}
\partial^2 \Psi_n +i [\partial_w\Phi,\partial_u\Psi_n]+i [\partial_w\Psi_n,\partial_u\Phi] = i [\partial^2\Phi+i [\partial_w\Phi,\partial_u\Phi],\Psi_{n-1}]=0
\,,
\label{eq:YMlinfromeom}
\end{eqnarray}
where the last step follows from the equation of motion for $\Phi$. The first equality above relies on the Jacobi identity. 
 \\
 
The discussion for self-dual gravity is analogous. For the Pleba\'nski equation, we can express the hierarchy of infinitesimal symmetries in terms of Hamiltonian vector fields $\{\Phi_{\theta_{n}}\}$ recursively via
\begin{align}
\partial_u\Phi_{\theta_{n}}=\partial_{\bar{w}}\Phi_{\theta_{n-1}}+[\partial_u\Phi_{\phi},\Phi_{\theta_{n-1}}],
\qquad 
\partial_w\Phi_{\theta_{n}}=\partial_{v}\Phi_{\theta_{n-1}}+[\partial_w\Phi_{\phi},\Phi_{\theta_{n-1}}].
\label{HierarchyPleba}
\end{align}
We can also write
\begin{align}
\Phi_{\partial_u\theta_{n}}=\Phi_{\partial_{\bar{w}}\theta_{n-1}}+\Phi_{\{\partial_u\phi,\Phi_{\theta_{n-1}}\}},
\qquad 
\Phi_{\partial_w\theta_{n}}=\Phi_{\partial_{v}\theta_{n-1}}+\Phi_{\{\partial_w\phi,\Phi_{\theta_{n-1}}\}}\,.
\end{align}
In analogy with the self-dual Yang-Mills case, we can take $\Phi_{\theta_{0}}$ as a constant Hamiltonian vector field, e.g., $\Phi_{\theta_{0}}=\partial_{u}-\partial_{w}$, so that $\Phi_{\theta_{1}}=[\Phi_{\phi},\Phi_{\theta_{0}}]=\Phi_{\{\phi,\theta_{0}\}}$, etc. For these two examples, we have $\theta_{0}=u+w$ and $\theta_{1}=\phi_{w}-\phi_{u}$, respectively. Then the conserved currents are obtained from
\begin{eqnarray}
J(x,\lambda)= \Phi_\theta ,
\qquad \text{such that} \quad [\partial_\mu, J^\mu]=0\,.
\end{eqnarray}
The analogous statement to \eqref{eq:YMlinfromeom} is
\begin{eqnarray}
\partial^2 \theta_n +\{\partial_w\phi,\partial_u\theta_n\}+\{\partial_w\theta_n,\partial_u\phi\} = \{\partial^2\phi+\{\partial_w\phi,\partial_u\phi\},\theta_{n-1}\}=0
\,,
\label{eq:gravlinfromeom}
\end{eqnarray}
or in terms of Hamiltonian vector fields
\begin{eqnarray}
\Phi_{\partial^2 \theta_n +\{\partial_w\phi,\partial_u\theta_n\}+i\{\partial_w\theta_n,\partial_u\phi\}} = [\Phi_{\partial^2\phi+\{\partial_w\phi,\partial_u\phi\}},\Phi_{\theta_{n-1}}]=0
\,.
\label{eq:gravlinfromeom2}
\end{eqnarray}



\section{Moyal deformations in gauge theory and gravity}
\label{sec:Moyal}

In the previous section, we reviewed the known double copy at the
level of equations of motion in self-dual Yang-Mills theory and
gravity, and also drew attention to integrability aspects that have
not been previously considered in a double copy context. Our overall
aim is to generalise this construction, and thus we shall proceed by
studying the kinds of generalisation of the above theories that are
possible. We will first look at {\it Moyal deformations} of our usual
self-dual theories, a concept which first arose in alternative
formulations of quantum mechanics
~\cite{Groenewold:1946kp,Baker:1958zza} as well as in the study of
field theories in non-commutative spacetimes
~\cite{doi:10.1142/5287,li2002physics}. The main idea is to replace
the standard product on functions with the {\it star product} $\star$,
defined in our present case as follows:
\begin{equation}
f\star g\equiv f\exp\left(\frac{i\hbar}{2}\buildrel{\leftrightarrow} \over {P}\right)g,
\label{starprod}
\end{equation}
where we have defined the operator:
\begin{align}
\buildrel{\leftrightarrow} \over {P} \; \equiv \;
\buildrel{\leftarrow} \over{\partial}_w
\buildrel{\rightarrow} \over{\partial}_u-
\buildrel{\leftarrow} \over{\partial}_u
\buildrel{\rightarrow} \over{\partial}_w
\label{PoissonOper}
\end{align}
that appears in the Poisson bracket of eq.~(\ref{Poissonbracket}):
\begin{equation}
\{f,g\}\equiv f \buildrel{\leftrightarrow} \over {P} g.
\label{Poissonbracket2}
\end{equation}
The star product is associative and non-commutative, and contains a
deformation parameter $\hbar$, whose notation stems from the original
context in (non-commutative) quantum theories, and whose
interpretation will be clarified below. Its use in defining deformed
equations of motion usually proceeds by defining the so-called {\it
  Moyal bracket}
\begin{equation}
\{f, g\}^{M}\equiv\frac{1}{i\hbar}(f\star g-g\star f),
\label{Moyalbracket}
\end{equation}
such that in the limit $\hbar\rightarrow 0$ we have
\begin{equation}
\lim_{\hbar\rightarrow 0}f\star g= fg, \qquad \lim_{\hbar\rightarrow 0}\{f, g\}^{M}=\{f, g\} .
\end{equation}
That is, the star product of two functions reduces to the conventional
(commutative) product of functions, and the Moyal bracket to the
Poisson bracket. By substituting eq.~(\ref{starprod}) into
eq.~(\ref{Moyalbracket}) and Taylor expanding in $\hbar$, we may write
a general expression for the Moyal bracket in terms of the Poisson
bracket, namely
\begin{eqnarray}
\{f,g\}^{M}&=&\sum_{s=0}^{\infty}\frac{(-1)^{s}\hbar^{2s}}{2^{2s}(2s+1)!}\sum
_{j=0}^{2s}(-1)^{j}
\binom{2s}{j}\{\partial_{w}^{2s-j}\partial_{u}^{j}f,\partial_{w}^{j}\partial_{u}^{2s-j}g\},
\label{MoyalPoisson}
\end{eqnarray}
or equivalently
\begin{eqnarray}
\{f,g\}^{M}=\sum_{s=0}^{\infty}\frac{(-1)^{s}\hbar^{2s}}{2^{2s}(2s+1)!}\sum _{j=0}^{2s+1}(-1)^{j} \binom{2s+1}{j}(\partial_{w}^{j}\partial_{u}^{2s+1-j}f)(\partial_{w}^{2s+1-j}\partial_{u}^{j}g).
\end{eqnarray}
In a similar fashion to the previous section, we can explore the
algebraic properties of this object. First, one can show that for any
three functions, the Moyal bracket satisfies the Jacobi identity
\begin{equation}
\{\{f,g\}^{M} ,h\}^{M}+\{\{g,h\}^{M},f\}^{M}+\{\{h,g\}^{M},f\}^{M}=0.
\label{MoyalJacobi}
\end{equation}
Then, we consider the particular relation
\begin{equation}
\{e^{-ip_{1}x},e^{-ip_{2}x}\}^{M}=-X^{M}(p_{1},p_{2})e^{-i(p_{1}+p_{2})x},
\label{expbracket}
\end{equation}
where the structure constant $X^{M}(p_{1},p_{2})$ is given by
\begin{eqnarray}
X^{M}(p_{1},p_{2})\equiv X (p_{1},p_{2})\sum_{s=0}^{\infty}\frac{(-1)^{s}\hbar^{2s}}{2^{2s}(2s+1)!}\sum _{j=0}^{2s}(-1)^{j} \binom{2s}{j}(p_{1w}p_{2u})^{2s-j}(p_{1u}p_{2w})^{j}.
\label{MoyalGenerators}
\end{eqnarray}
This is a generalisation of the quantity $X(p_1,p_2)$ encountered in
eq.~(\ref{Fp1p2def}), and reduces smoothly to the latter in the
appropriate limit:
\begin{equation}
\lim_{\hbar\rightarrow 0}X^{M}(p_{1},p_{2})=X (p_{1},p_{2}).
\label{XMlim}
\end{equation}
Furthermore, using eqs.~(\ref{MoyalJacobi}) and
(\ref{MoyalGenerators}), it is easy to see that the structure constant
$X^{M}$ also satisfies the Jacobi identity
\begin{equation}
X^{M}(p_{1},p_{2})X^{M}(p_{1}+p_{2},p_{3})+X^{M}(p_{2},p_{3})X^{M}(p_{2}+p_{3},p_{1})+X^{M}(p_{3},p_{1})X^{M}(p_{3}+p_{1},p_{2})=0.
\label{JacobiXM}
\end{equation}
We can therefore use the deformed structure constant $X^M(p_1,p_2)$ as
a building block in constructing generalised field theories that obey
double copy relations, and will see a number of examples below.\\

%

Another useful realisation is the Moyal deformation of the Lie bracket. Naively, we could define this as
\begin{equation}
[V,W]^{M}\equiv \sum_{s=0}^{\infty}\frac{(-1)^{s}\hbar^{2s}}{2^{2s}(2s+1)!}\sum _{j=0}^{2s}(-1)^{j} \binom{2s}{j}[\partial_{w}^{2s-j}\partial_{u}^{j}V,\partial_{w}^{j}\partial_{u}^{2s-j}W].
\label{MoyalComm}
\end{equation}
For vector fields $V$ and $W$, the derivatives inside the Lie bracket are understood as acting as $\partial_uV\equiv[\partial_u,V]$, etc. Generically, this idea is naive, because such a Moyal bracket does not in general satisfy the Jacobi identity. A particular case where it does indeed satisfy it is that of Hamiltonian vector fields, 
\begin{eqnarray}
[\Phi_{f},\Phi_{g}]^{M}&=&\sum_{s=0}^{\infty}\frac{(-1)^{s}\hbar^{2s}}{2^{2s}(2s+1)!}\sum _{j=0}^{2s}(-1)^{j} \binom{2s}{j}[\partial_{w}^{2s-j}\partial_{u}^{j}\Phi_{f},\partial_{w}^{j}\partial_{u}^{2s-j}\Phi_{g}]\nonumber\\
&=&\sum_{s=0}^{\infty}\frac{(-1)^{s}\hbar^{2s}}{2^{2s}(2s+1)!}\sum _{j=0}^{2s}(-1)^{j} \binom{2s}{j}\Phi_{\{\partial_{w}^{2s-j}\partial_{u}^{j}f,\partial_{w}^{j}\partial_{u}^{2s-j}g\}}\nonumber\\
&=&\Phi_{\{f,g\}^{M}},
\label{eq:LieMPoiM}
\end{eqnarray}
such that, analogously to the Jacobi identity (\ref{MoyalJacobi}),
\begin{equation}
[[\Phi_{f},\Phi_{g}]^{M},\Phi_{h}]^{M}+[[\Phi_{g},\Phi_{h}]^{M},\Phi_{f}]^{M}+[[\Phi_{h},\Phi_{f}]^{M},\Phi_{g}]^{M}=0.
\label{Jacobi:DeformedLieBracket}
\end{equation}

\subsection{Deformed self-dual gravity}

As a first application of the above ideas, we will consider a Moyal
deformation of the Pleba\'nski equation, which was investigated in
\cite{Plebanski:1995gk,Plebanski:1996np,TAKASAKI1994332}. 
The deformed Pleba\'nski
(second heavenly) equation is
\begin{equation}
\partial^{2}\phi+\{\partial_{w}\phi,\partial_{u}\phi\}^{M}=0.
\label{MHK}
\end{equation}
In momentum space, this deformed equation can be expressed as 
\begin{equation}
\phi(k)=\frac{1}{2} \int \dd p_{1}\dd p_{2}\frac{\del (p_{1}+p_{2}-k)}{k^{2}}X^{M}(p_{1},p_{2})X (p_{1},p_{2})\phi(p_{1})\phi(p_{2}),
\label{MoyalMomentum}
\end{equation}
and it is clear from eq. (\ref{XMlim}) that the usual Pleba\'nski
equation both in coordinate space (\ref{Plebanski}) and in momentum
space eq. (\ref{Gravsol}) are recovered in the limit $\hbar\rightarrow
0$.\\

The form of eq. (\ref{MoyalMomentum}) is particularly appealing from a
double copy point of view. That is, all of the theories considered in
this paper have momentum-space integral equations whose kernels involve products of
structure constants, and the present case can be obtained by a process
analogous to the double copy of eq. (\ref{DoubleCopy}), making instead
the replacement
\begin{equation}
f^{abc}\rightarrow X^{M}(p_1,p_2),
\end{equation}
in eq.~(\ref{YMsol}). Below, we will extend this philosophy to obtain
a web of theories with certain desirable properties. Before doing so,
however, it is interesting to examine the integrability of the theory
of eq.~(\ref{MHK}). This has been investigated in
\cite{Strachan:1996gx,TAKASAKI1994332}, where as usual one may address
this question by formulating a Lax pair. There are two possible ways
to do this, namely we may consider a Lax pair consisting of undeformed vectors satisfying a deformed bracket, or `deformed vectors' satisfying an undeformed bracket. The two options are related to attributing $X(p_1,p_2)$ or $X^M(p_1,p_2)$ in eq.~\eqref{MoyalMomentum} to the Lax compatibility condition or to the Lax pair. For the
undeformed Lax pair, we can take the results of eq. (\ref{Laxgrav})
and use \eqref{eq:LieMPoiM} to show that they satisfy the deformed compatibility condition
\begin{equation}
[L,M]^{M}=0
\label{NonDeformed}
\end{equation}
if $\phi$ satisfies the deformed Pleba\'nski eq. (\ref{MHK}).\\

The argument when using a deformed pair (but undeformed bracket) is
more involved. First, let us explain the deformation of the vectors as
follows.  We can rewrite the Hamiltonian vector field $\Phi_{f}$ in
the $(w,u)$ plane, given in eq. (\ref{Poisson}), using the Poisson operator
(\ref{PoissonOper}) in the form
\begin{equation}
\Phi_{f}\equiv f\buildrel{\leftrightarrow} \over {P}=f_w \partial_{u}- f_u \partial_{w}.
\end{equation}
Clearly, this vector satisfies the relation
\begin{equation}
\Phi_f (g)=f\buildrel{\leftrightarrow} \over {P} g=\{f,g\}.
\label{HamilPoisson}
\end{equation}
Now, following the Poisson bracket case, eq. (\ref{HamilPoisson}), we can define (in a slight abuse of language) a \textit{deformed Hamiltonian vector field}, $\Phi^{M}_{f}$, by 
\begin{equation}
\Phi^{M}_{f}(g)\equiv\{f,g\}^{M}.
\end{equation}
Then we have \cite{Strachan:1996gx,STRACHAN199263}
\begin{align}
\Phi^{M}_{f}&= \frac{2}{\hbar}f\sin\left(\frac{\hbar}{2}\buildrel{\leftrightarrow} \over {P}\right)\no\\
&=\sum_{s=0}^{\infty}\frac{(-1)^{s}\hbar^{2s}}{2^{2s}(2s+1)!}\sum _{j=0}^{2s+1}(-1)^{j} \binom{2s+1}{j}(\partial_{w}^{j}\partial_{u}^{2s+1-j}f)\partial_{w}^{2s+1-j}\partial_{u}^{j}.
\end{align}
Note that in the $\hbar\rightarrow 0$ limit, we recover a conventional
Hamiltonian vector field, i.e. $\lim_{\hbar\rightarrow
  0}\Phi^{M}_{f}=\Phi_{f}$. It follows from the above that the Lie
bracket of deformed Hamiltonian vector fields is
\begin{equation}
[\Phi^{M}_{f},\Phi^{M}_{g}]=\Phi^{M}_{\{f,g\}^{M}}.
\end{equation}
Note that, following the Jacobi identity (\ref{MoyalJacobi}), the deformed Hamiltonian vector fields satisfies the Jacobi identity 
\begin{equation}
[[\Phi^{M}_{f},\Phi^{M}_{g}],\Phi^{M}_{h}]+[[\Phi^{M}_{g},\Phi^{M}_{h}],\Phi^{M}_{f}]+[[\Phi^{M}_{h},\Phi^{M}_{f}],\Phi^{M}_{g}]=0.
\label{Jacobi:DeformedHamiltonianVectors}
\end{equation}
Taking the `deformed vectors'
\begin{equation}
L^{M}=\partial_{u}-\lambda\left(\partial_{\bar{w}}+\Phi_{\phi_{u}}^{M}\right),\qquad\qquad M^{M}=\partial_{w}-\lambda\left(\partial_{v}+\Phi_{\phi_{w}}^{M}\right),
\label{DefoLaxHK}
\end{equation}
we can check that they satisfy the condition
\begin{equation}
[L^{M},M^{M}]=0
\label{DefoLaxHK2}
\end{equation}
if $\phi$
satisfies the deformed Pleba\'nski eq.~(\ref{MHK}). To see this, we
note that only the $\lambda^{2}$ term does not vanish trivially,
taking the form
\begin{eqnarray}
[\partial_{v}+\Phi_{\phi_{w}}^{M},\partial_{\bar{w}}+\Phi_{\phi_{u}}^{M}]&=& \partial^{2}\Phi_{\phi}^{M}+[\Phi_{\phi_{w}}^{M},\Phi_{\phi_{u}}^{M}]\nonumber\\
&=&\Phi_{\partial^{2}\phi+\{\phi_{w},\phi_{u}\}^{M}}^{M}\nonumber\\
&=&0.
\end{eqnarray}
In this case we have the deformed vectors
\begin{align}
\nonumber U&=\partial_u, & V&=\partial_v+\Phi_{\phi_{w}}^{M},\\
W&=\partial_w, & \bar{W}&=\partial_{\bar{w}}+\Phi_{\phi_{u}}^{M}.
\label{PleVecDefor}
\end{align}
The two equivalent approaches mentioned above, of deforming either the bracket or the Lax pair, are represented by \eqref{NonDeformed} and \eqref{DefoLaxHK2}.\\

We now proceed in analogy to section \ref{sec:Tower} to obtain an infinite tower of conserved currents for the deformed theory. First, the linearised equation of motion is obtained from the deformed Pleba\'nski equation (\ref{MHK}) via the replacement $\Phi^{M}_{\phi}\rightarrow \Phi^{M}_{\phi}+\Phi^{M}_{\theta}$, is
\begin{equation}
\partial^{2}\Phi^{M}_{\theta}+[\partial_{w}\Phi^{M}_{\phi},\partial_{u}\Phi^{M}_{\theta}]+[\partial_{w}\Phi^{M}_{\theta},\partial_{u}\Phi^{M}_{\phi}]=0.
\label{linearDefPleba}
\end{equation}
Now, let us consider $\Phi^{M}_{\theta}$ of the form
\begin{equation}
\Phi^{M}_{\theta}=\sum^{\infty}_{n=0}\Phi^{M}_{\theta_{n}}\lambda^{n}.
\end{equation}
We can express the infinite tower of linearised solutions $\{\Phi^{M}_{\theta_{n}}\}$ to eq. (\ref{linearDefPleba}) recursively via
\begin{align}
\partial_u\Phi^{M}_{\theta_{n}}=\partial_{\bar{w}}\Phi^{M}_{\theta_{n-1}}+[\partial_u\Phi^{M}_{\phi},\Phi^{M}_{\theta_{n-1}}],
\qquad 
\partial_w\Phi^{M}_{\theta_{n}}=\partial_{v}\Phi^{M}_{\theta_{n-1}}+[\partial_w\Phi^{M}_{\phi},\Phi^{M}_{\theta_{n-1}}],
\label{HierarchyDefPleba1}
\end{align}
with $\Phi^{M}_{\theta_{-1}}\equiv0$. This $\Phi^{M}_{\theta_{n}}$ satisfies the compatibility condition $\partial_{u}\partial_{w}\Phi^{M}_{\theta_{n}}=\partial_{w}\partial_{u}\Phi^{M}_{\theta_{n}}$ and it  is a solution of the linearised equation (\ref{linearDefPleba}), using the Jacobi identity (\ref{Jacobi:DeformedHamiltonianVectors}). Alternatively, the same infinite tower of linearised solutions can be obtained from Hamiltonian vector fields but using the deformed bracket. In this notation, we have the recursive equations
\begin{align}
\partial_u\Phi_{\theta_{n}}=\partial_{\bar{w}}\Phi_{\theta_{n-1}}+[\partial_u\Phi_{\phi},\Phi_{\theta_{n-1}}]^{M},
\qquad 
\partial_w\Phi_{\theta_{n}}=\partial_{v}\Phi_{\theta_{n-1}}+[\partial_w\Phi_{\phi},\Phi_{\theta_{n-1}}]^{M},
\label{HierarchyDefPleba2}
\end{align}
with $\Phi_{\theta_{-1}}\equiv0$. As in the deformed Hamiltonian vector fields, this $\Phi_{\theta_{n}}$ satisfies the compatibility condition $\partial_{u}\partial_{w}\Phi_{\theta_{n}}=\partial_{w}\partial_{u}\Phi_{\theta_{n}}$ and it  is a solution of the linearised equation 
\begin{equation}
\partial^{2}\Phi_{\theta}+[\partial_{w}\Phi_{\phi},\partial_{u}\Phi_{\theta}]^{M}+[\partial_{w}\Phi_{\theta},\partial_{u}\Phi_{\phi}]^{M}=0.
\end{equation}
using the Jacobi identity (\ref{Jacobi:DeformedLieBracket}).\\

The Moyal-deformed self-dual gravity shares features and properties with the undeformed version. One such property, as discussed above, is integrability. In terms of the language of the first heavenly equation, eq.~(\ref{FirstHeavenly}), this property can be expressed in a concise geometric way with an associated 2-form $\mathbf{\Omega}$, which satisfies the equations $d\mathbf{\Omega}=0$ and $\mathbf{\Omega}\wedge\mathbf{\Omega}=0$ \cite{Strachan:1996gx}. This is the K\"ahler form, which is related to the K\"ahler potential $\Omega$ as $\mathbf{\Omega}=\Omega_{\mu\nu}dx^{\mu}\wedge d\tilde{x}^{\nu}$, where $x^{\mu}=\{w,u\}$ and $\tilde{x}^{\nu}=\{y,z\}$. In \cite{Maceda_2019}, a Moyal deformation of the first heavenly equation was considered, and a four-dimensional Moyal deformed integrable K\"ahler manifold was constructed by imposing in a consistent way that the deformed 2-form $\mathbf{\Omega}$ is closed, Hermitian and has unit determinant. This procedure provides an alternative but equivalent approach to equation \eqref{MHK}, based on the second heavenly equation.

\subsubsection{Deformed self-dual gravity as undeformed self-dual Yang-Mills}
\label{sec:DSDGvsSDG}

In this subsection, we take an alternative approach to the Moyal deformation of self-dual gravity, which leads directly to undeformed self-dual Yang-Mills. Instead of the deformation $X(p_{1},p_{2})\to X^{M}(p_{1},p_{2})$ considered above, we will see that we can think of $X(p_{1},p_{2})\to f^{abc}$ as a Moyal deformation, at least in a particular construction. \\

 In order to do that we follow now Ref. \cite{doi:10.1063/1.528788}, which considered the case of SDYM in $\mathbb{R}^{1,D-1}$ with su$(N)$
or sl$(N,\mathbb{C})$ Lie algebra valued fields.\footnote{By SDYM beyond $D=4$, we mean simply the straightforward extension of \eqref{YMEOM}, where the wave operator is $D$-dimensional. The dimensionality is not relevant for our purposes.}
In Ref. \cite{doi:10.1063/1.528788}, a novel version of the basis for
these algebras was found, in terms of a double index notation
$m=(m_1,m_2)$.  In more detail, the Lie algebra generators can be
written as
\begin{equation}
L_{m} \equiv -{N \over 2 \pi} \omega^{m_1m_2 \over 2} S^{m_1} T^{m_2},
\label{TrigonometricAlgebra}
\end{equation}
where $\omega = \exp\big({2 \pi i \over N} \big)$, and  $S$ and $T$ are $N \times N$ matrices such that $S^N =T^N=-1$ and 
$T \cdot S = \omega\, S \cdot T$. These generators satisfy
\begin{equation}
[L_{m},L_{n}]=i{\cal F}_{m,n}^{(m+n)}L_{m + n},
\label{algebram}
\end{equation}
where ${\cal F}_{m,n}^{(m+ n)} = {N \over \pi} \sin \big({\pi \over N} m \times n \big)$. Thus, the analogue of eq. (\ref{YMsol}) is given by
\begin{eqnarray}
\Phi^{m + n}(k)
= \frac{1}{2}\int \dd p_{1}\dd p_{2}\frac{\del (p_{1}+p_{2}-k)}{k^{2}}X(p_{1},p_{2}) {N \over \pi} \sin \bigg({\pi \over N} m \times n \bigg) \Phi^{m}(p_{1})\Phi^{n}(p_{2}).
\label{DSDYMMomentumnew}
\end{eqnarray}
In this basis, scalars $\Phi$ are expanded as $\Phi(x) = \sum_{m} \Phi^{m}(x) L_{m}$. \\

In the limit $N \to \infty$ for the Lie algebra in eq. (\ref{algebram}), we get the sdiff$(T^2)$ algebra
\begin{equation}
\{e_{m},e_{n}\}_{T} = (m \times n) e_{m + n},
\end{equation} 
where $m \times n=m_{1}n_{2}-m_{2}n_{1}$, and the Poisson bracket on $T^{2}$ is $\{f,g\}_{T}\equiv f_{x_{2}}g_{x_{1}}-f_{x_{1}}g_{x_{2}}$.
The generators of sdiff($T^2$) with local coordinates $(x_{1},x_{2})$ are
\begin{equation}
e_m \equiv \exp i(m_1 x_{1} + m_2 x_{2}).
\end{equation} 

Furthermore there is an isomorphism which maps the Lie bracket (\ref{algebram}) into the Moyal bracket for  $e_{m}$  :
\begin{equation}
\{e_{m}, e_{n}\}_{T}^M = {2 \over \hbar} \sin \bigg({\hbar \over 2} m \times n  \bigg) e_{m + n},
\end{equation}
where we have made the identification
\begin{equation}
\hbar =\frac{2 \pi }{ N}.
\end{equation} 
Thus the large-$N$ limit  corresponds to the limit $\hbar \to 0$.\\

Then, the equation of motion on  $\mathbb{R}^{1,D-3} \times T^2$ describes an integrable Moyal deformation of self-dual gravity equation known as the $\star$-SDYM system \cite{Forma_ski_20051,Forma_ski_20052} given by
\begin{equation}
\partial^{2}\phi+\{\partial_{w}\phi,\partial_{u}\phi\}_{T}^{M}=0,
\label{DSDYMdos}
\end{equation}
for $\phi = \sum_{m}\phi^{m}(\hbar;\vec{x},x_{1},x_{2}) e_{m}$ 
and $\phi^{m}= \sum_{n=0}^\infty \hbar^n
\phi^{m}_n$. Equation~(\ref{DSDYMdos}) can then be written as
\begin{eqnarray}
\phi^{m + n}(k)
={1 \over \hbar}\int \dd p_{1}\dd p_{2}\frac{\del (p_{1}+p_{2}-k)}{k^{2}}X(p_{1},p_{2}) 
 \sin \bigg({\hbar \over 2} m \times n  \bigg) \phi^{m}(p_{1})\phi^{n}(p_{2}).
\label{DSDYMMomentumdos}
\end{eqnarray}
The {\it master equation} (\ref{DSDYMdos}) is proven
to correspond to an {\it integrable} system \cite{Forma_ski_20051},
by showing the existence of Lax pairs, an infinite
hierarchy of conserved quantities, and a twistor
construction. Moreover in Ref. \cite{Forma_ski_20052} some explicit
examples are given. The direct map between \eqref{DSDYMMomentumdos} and the SDYM equation \eqref{DSDYMMomentumnew} is clear.\\

Taking the limit $\hbar \to 0$ in eq. (\ref{DSDYMMomentumdos}), we get the Pleba\'nski-Przanowski equation \cite{Plebanski:1996np} 
\begin{eqnarray}
\phi^{m + n}_{0}(k)
={1 \over 2}\int \dd p_{1}\dd p_{2}\frac{\del (p_{1}+p_{2}-k)}{k^{2}}X(p_{1},p_{2}) 
(m \times n) \phi^{m}_{0}(p_{1})\phi^{n}_{0}(p_{2}).
\label{DSDYMMomentumtres}
\end{eqnarray}
Equivalently, one can take the limit $N\rightarrow\infty$ in eq. (\ref{DSDYMMomentumnew}) on $\mathbb{R}^{1,D-3}$. For instance, the $D =4$ case corresponds to $\mathbb{R}^{1,1} \times T^2$. In this case the gauge theory is the principal chiral model, which is an integrable two-dimensional reduction of SDYM equations on $\mathbb{R}^{1,3}$. This model leads to the Husain-Park heavenly equation which was discussed in \cite{doi:10.1142/S0217732398003405,Forma_ski_20052}.  \\

It is worth mentioning that this procedure can be employed to interpolate between all theories shown in figure 1.  For instance, we can start from the biadjoint scalar theory and take in eq. (\ref{BAdjsol}) two times the representation of the Lie algebra  in the trigonometric basis \cite{doi:10.1063/1.528788} i.e. $f^{abc} \to {2 \over \hbar}  \sin \big({\hbar \over 2} m \times n \big)$ and  
$\widetilde{f}^{a'b'c'} \to {2 \over \hbar '}  \sin \big({\hbar ' \over 2} m' \times n' \big)$. Taking the limit, for instance, $\hbar \to 0$ this represents the inverse zeroth copy and we get self-dual Yang-Mills equation (\ref{YMsol}). A further limit 
$\hbar ' \to 0$ leads straightforwardly to the double copy, i.e.~the self-dual gravity equation (\ref{Gravsol}). It is also possible to start from the self-dual gravity equation (\ref{Gravsol}) and take its single copy as a Moyal deformation of gravity with $\hbar '\not =0$. A further Moyal deformation with $\hbar \not =0$ in the gauge theory leads directly to the zeroth copy giving the biadjoint
scalar theory. Thus, this procedure allows us to interpolate with two continuous parameters $\hbar$ and $\hbar '$ among the set of theories mentioned in figure 1. Of course, this particular construction relied on periodic identifications of coordinates, so that a $T^2$ arises, and the Fourier modes are discrete. If they are not discrete, then we will have structure constants labeled by continuous indices like $X^M(p_1,p_2)$ in the earlier discussion.

\subsection{Deformed Self-Dual Yang-Mills}
\label{sec:YMdeform}

We will now explore the Moyal deformation of self-dual Yang-Mills theory. We will see, however, that the deformation that fits in with the colour-kinematics duality does not coincide with the most commonly considered Moyal deformation of self-dual Yang-Mills -- whereas the latter deformation is integrable \cite{TAKASAKI2001291,HAMANAKA2006368}, the former does not preserve it. \\

%
%

Starting from the deformed Pleba\'nski
equation for self-dual gravity in momentum space (\ref{MoyalMomentum}), we have two
different single copies. One of them, replacing
$X^{M}(p_{1},p_{2})\rightarrow f^{abc}$, is the usual self-dual
Yang-Mills theory. On the other hand, we can take
$X(p_{1},p_{2})\rightarrow f^{abc}$ and we obtain a deformed version
of self-dual Yang-Mills. The equation of motion for this gauge theory
in momentum space is
\begin{eqnarray}
\Phi^{a}(k)
=\frac{1}{2}\int \dd p_{1}\dd p_{2}\frac{\del (p_{1}+p_{2}-k)}{k^{2}}X^{M}(p_{1},p_{2})f^{abc}\Phi^{b}(p_{1})\Phi^{c}(p_{2}).
\label{DSDYMMomentum}
\end{eqnarray}
In the limit $\hbar\rightarrow 0$, we recover the usual SDYM equation
(\ref{YMsol}). Furthermore, one may show 
that eq.~(\ref{DSDYMMomentum}) is
equivalent to the position space equation
\begin{equation}
\partial^2\Phi+i[\partial_w \Phi,\partial_u \Phi]^{M}=0,
\label{DSDYMPosition}
\end{equation}
where we are using the Moyal bracket (\ref{MoyalComm}). This theory differs from the usual Moyal deformation of self-dual
Yang-Mills theory that is considered in the literature, which can be
written as
\begin{equation}
\partial^2\Phi+i[\partial_w \Phi,\partial_u \Phi]_\star=0,\qquad
[A,B]_\star\equiv A\star B-B\star A.
\label{YMnoncomm}
\end{equation}
This latter equation has a very specific interpretation: it is the
theory one obtains upon introducing self-dual Yang-Mills theory in a
non-commutative spacetime. To our knowledge, no such interpretation exists for
eq.~(\ref{DSDYMPosition}), which is the unique theory one obtains upon
single copying the deformed Pleba\'{n}ski equation in the
above-mentioned fashion. Both eq.~(\ref{DSDYMPosition}) and
eq.~(\ref{YMnoncomm}) reduce to the usual self-dual Yang-Mills theory
in the limit $\hbar\rightarrow 0$. The relation between these two distinct deformations is not obvious. In fact, the colour structure of eq.~\eqref{YMnoncomm} is very different from the one typically found in gauge theories. For instance, a calculation shows that
\begin{eqnarray}
i[\partial_w (\Phi^aT^a),\partial_u (\Phi^bT^b)]_\star = &
-\frac12 f^{abc}T^c\,\frac{d}{d\hbar} \left(\hbar\,\{ \Phi^a, \Phi^b\}^M \right)
-\hbar\, T^{(a}T^{b)}\, \{\partial_w \Phi^a,\partial_u \Phi^b\}^M \,
 \nonumber\\
= &
-\frac12 f^{abc}T^c\,\{ \Phi^a, \Phi^b\}
-\hbar\, T^{(a}T^{b)}\, \{\partial_w \Phi^a,\partial_u \Phi^b\} + {\mathcal O}(\hbar^2) .
\label{eq:moyalrel}
\end{eqnarray} 
In the right-hand side of
the first line, the first/second term has only odd/even powers of
$\hbar$. We can see right away that $\Phi$ is generically not a colour
Lie algebra element in the usual Moyal deformation (\ref{YMnoncomm}),
i.e., $\Phi\neq\Phi^aT^a$; because if you start with $\Phi=\Phi^aT^a$ in the left-hand side of eq.~\eqref{YMnoncomm}, a distinct colour structure is generated, $T^{(a}T^{b)}$, which is not an element of the Lie algebra. Instead, $\Phi$ lives in the enveloping
algebra generated by elements $T^{(a_1}T^{a_2}\cdots T^{a_n)}$. This
contrasts with the single copy theory case of eq.~(\ref{DSDYMPosition}), where $\Phi$ lives in the
Lie algebra. The question of whether there is a more direct physical
interpretation of eq.~(\ref{DSDYMPosition}) is interesting.\footnote{There are several instances in the double copy literature where one theory of no obvious relevance of its own is very useful for providing building blocks, via the double copy, for more physically relevant theories.} For our present purposes, we shall simply examine the question of integrability, by attempting to construct a Lax pair following an
analogous process to that of SDYM. We will take a naive approach in order to see where it fails. Applying the Moyal
bracket (\ref{MoyalComm}) for a gauge field, one may first look at
the components of the deformed field strength tensor
\begin{equation}
F_{\mu\nu}^{M}\equiv \partial_{\mu}A_{\nu}-\partial_{\nu}A_{\mu}-i[A_{\mu},A_{\nu}]^{M} ,
\end{equation}
and the deformed self-duality conditions take the form
\begin{gather}
F_{uw}^{M}=0,\nonumber\\
F_{v\bar{w}}^{M}=0,\nonumber\\
F_{uv}^{M}-F_{w\bar{w}}^{M}=0.
\end{gather}
These equations arise as the condition $[L,M]^{M} =0$ for the two
operators defined in eq. (\ref{LaxGaugeRedSD}). However, this does not
by itself guarantee integrability, which may be seen upon trying to
construct a hierarchy of linearised solutions. The replacement $\Phi\rightarrow \Phi+\Psi$ results in the linearised equation
\begin{equation}
\partial^2\Psi+i[\partial_w \Psi,\partial_u \Phi]^{M}+i[\partial_w \Phi,\partial_u \Psi]^{M}=0.
\label{LinearesedDefSDYM}
\end{equation}
Expanding $\Psi$ in powers of the spectral parameter $\lambda$, we can express the infinite tower $\{\Psi_{n}\}$ recursively via
\begin{align}
\partial_u\Psi_n=\partial_{\bar{w}}\Psi_{n-1}+i[\partial_u\Phi,\Psi_{n-1}]^{M},
\qquad 
\partial_w\Psi_n=\partial_{v}\Psi_{n-1}+i[\partial_w\Phi,\Psi_{n-1}]^{M},
\label{HierarchyDefSDYM}
\end{align}
with $\Psi_{-1}\equiv0$. 
One may be tempted to conclude that the theory is integrable. However, this type of construction relies crucially on the Jacobi identity for the bracket, which fails in the present case of the Moyal bracket (\ref{MoyalComm}).\footnote{We made comments regarding this property near (\ref{MoyalComm}), where we noted that, when the Moyal-deformed bracket was applied to vector fields, the Jacobi identity did not apply in general, but did apply to Hamiltonian vector fields.} It relies on the Jacobi identity in order to prove that this infinite tower is a solution of the linearised equation (\ref{LinearesedDefSDYM}) -- without the Jacobi identity, one cannot show that $\Psi_{n}$ is a symmetry if $\Psi_{n-1}$ is. Even before that, the Jacobi identity is crucial in order to relate the linearised equation to the presumptive Lax pair. To see in detail how the Jacobi identity fails, let us take three gauge fields $V=V^{a}T^{a}$, $W=W^{b}T^{b}$ and $U=U^{c}T^{c}$. First, the Moyal bracket of $V$ and $W$ is explicitly 
\begin{eqnarray}
[V,W]^{M}&=&i \sum_{s=0}^{\infty}\frac{(-1)^{s}\hbar^{2s}}{2^{2s}(2s+1)!}\sum _{j=0}^{2s}(-1)^{j} \binom{2s}{j}f^{abd}(\partial_{w}^{2s-j}\partial_{u}^{j}V^{a})(\partial_{w}^{j}\partial_{u}^{2s-j}W^{b})T^{d}\nonumber\\
&=&i \sum_{s=0}^{\infty}\frac{(-1)^{s}\hbar^{2s}}{2^{2s}(2s+1)!}\sum _{j=0}^{2s}(-1)^{j} \binom{2s}{j}f^{abd}V^{a}_{2s-j,j}W^{b}_{j,2s-j}T^{d}\nonumber\\
&=&if^{abd}V^{a}W^{b}T^{d}+ {\mathcal O}(\hbar^2),
\end{eqnarray}
using the notation $\partial_{w}^{n}\partial_{u}^{m}F=F_{n,m}$. Now, for three gauge fields, we have
\begin{multline}
[[V,W]^{M},U]^{M}=-f^{abd}f^{dce}T^{e}\\
\times \sum_{s,s'=0}^{\infty}\frac{(-1)^{s+s'}\hbar^{2(s+s')}}{2^{2(s+s')}(2s+1)!(2s'+1)!}\sum _{j=0}^{2s}\sum _{j'=0}^{2s'}(-1)^{j+j'} \binom{2s}{j}\binom{2s'}{j'}(V^{a}_{2s-j,j}W^{b}_{j,2s-j})_{2s'-j',j'}U^{c}_{j',2s'-j'}\\
=-f^{abd}f^{dce}T^{e}\left[V^{a}W^{b}U^{c}-\frac{\hbar^{2}}{24}\sum _{j=0}^{2}(-1)^{j} \binom{2}{j}\left(V^{a}_{2-j,j}W^{b}_{j,2-j}U^{c}+(V^{a}W^{b})_{2-j,j}U^{c}_{j,2-j}\right)\right] + {\mathcal O}(\hbar^4).
\end{multline}
Finally, the Jacobi type equation gives
\begin{multline}
[[V,W]^{M},U]^{M}+[[W,U]^{M},V]^{M}+[[U,V]^{M},W]^{M}\\
=-(f^{abd}f^{dce}+f^{bcd}f^{dae}+f^{cad}f^{dbe})T^{e}V^{a}W^{b}U^{c}+ {\mathcal O}(\hbar^2).
\end{multline}
The leading order term, ${\mathcal O}(\hbar^0)$, vanishes due to the colour Jacobi identity. However, the first subleading term is proportional to
\begin{multline}
\sum _{j=0}^{2}(-1)^{j} \binom{2}{j}\left[f^{abd}f^{dce}\left(V^{a}_{2-j,j}W^{b}_{j,2-j}U^{c}+(V^{a}W^{b})_{2-j,j}U^{c}_{j,2-j}\right)\right.\\
\left.+f^{bcd}f^{dae}\left(W^{b}_{2-j,j}U^{c}_{j,2-j}V^{a}+(W^{b}U^{c})_{2-j,j}V^{a}_{j,2-j}\right)+f^{cad}f^{dbe}\left(U^{c}_{2-j,j}V^{a}_{j,2-j}W^{b}+(U^{c}V^{a})_{2-j,j}W^{b}_{j,2-j}\right)\right]\\
\neq 0,
\end{multline}
that is, it does not vanish generically. Hence, $[[V,W]^{M},U]^{M}+[[W,U]^{M},V]^{M}+[[U,V]^{M},W]^{M}\neq 0$. The integrability property relies implicitly on the Jacobi identity for the appropriate bracket, which fails in this case. Therefore, the deformation \eqref{DSDYMPosition} of the SDYM equation of motion, which we arrived at by taking the single copy of Moyal-deformed SDG that keeps the factor $X^{M}(p_{1},p_{2})$, and changes $X(p_{1},p_{2})\rightarrow f^{abc}$, is not integrable. This is because the integrability of Moyal-deformed SDG relies precisely on the factor $X(p_{1},p_{2})$, which we discarded here.

\subsection{Doubly deformed self-dual gravity}
\label{sec:doubledeform}

We have so far been constructing various theories from known self-dual
integral equations, by replacing different structure constants by
their deformed counterparts. Continuing this procedure, we can
consider the theory one obtains by taking two copies of the
Moyal-deformed kinematic algebra, whose structure constant is
$X^M(p_1,p_2)$. This can be obtained e.g. by taking the deformed
self-dual Yang-Mills theory of eq.~(\ref{DSDYMMomentum}), and
replacing $f^{abc}\rightarrow X^{M'}(p_{1},p_{2})$. The resulting
equation of motion in momentum space then takes the form
\begin{equation}
\phi(k)=\frac{1}{2}\int \dd p_{1}\dd p_{2}\frac{\del (p_{1}+p_{2}-k)}{k^{2}}X ^{M}(p_{1},p_{2})X^{M'}(p_{1},p_{2})\phi(p_{1})\phi(p_{2}),
\end{equation}
where
\begin{equation}
X^{M'}(p_{1},p_{2})=X (p_{1},p_{2})\sum_{s=0}^{\infty}\frac{(-1)^{s}\hbar'^{2s}}{2^{2s}(2s+1)!}\sum _{j=0}^{2s}(-1)^{j} \binom{2s}{j}(p_{1w}p_{2u})^{2s-j}(p_{1u}p_{2w})^{j},
\end{equation}
and we have introduced a second deformation parameter $\hbar'$ to be
associated with $M'$. In position space this equation is
\begin{equation}
\partial^{2}\phi+\sum_{s=0}^{\infty}\frac{(-1)^{s}\hbar^{2s}}{2^{2s}(2s+1)!}\sum _{j=0}^{2s}(-1)^{j} \binom{2s}{j}\{\partial_{w}^{2s-j+1}\partial_{u}^{j}\phi,\partial_{w}^{j}\partial_{u}^{2s-j+1}\phi\}^{M'}=0.
\label{DDPosition}
\end{equation}
If we take either of the limits $\hbar\rightarrow 0$ or
$\hbar'\rightarrow 0$, we recover the deformed Pleba\'nski equation
(\ref{MHK}). Furthermore, it is straightforward to verify that
eq.~(\ref{DDPosition}) arises from the (doubly deformed) Lax Pair
condition
\begin{equation}
[L^{M'},M^{M'}]^{M}=0.
\end{equation}
To see this, one may first note that only the $\lambda^{2}$ does not
vanish trivially, which in turn leads explicitly to
\begin{eqnarray}
[\partial_{v}+\Phi_{\phi_{w}}^{M'},\partial_{\bar{w}}+\Phi_{\phi_{u}}^{M'}]^{M}&=& \partial^2\, \Phi^{M'}_{\phi}+[\Phi^{M'}_{\phi_{w}},\Phi^{M'}_{\phi_{u}}]^{M} \no\\
&=& \Phi^{M'}_{\partial^{2}\phi+\sum_{s=0}^{\infty}\frac{(-1)^{s}\hbar^{2s}}{2^{2s}(2s+1)!}\sum _{j=0}^{2s}(-1)^{j} \binom{2s}{j}\{\partial_{w}^{2s+1-j}\partial_{u}^{j}\phi,\partial_{w}^{j}\partial_{u}^{2s+1-j}\phi\}^{M}}\no\\
&=&0.
\end{eqnarray}
Moreover, we can obtain the same equation with
$[L^{M},M^{M}]^{M'}=0$. For this double deformed Pleba\'nski equation
(\ref{DDPosition}), we obtain a linearised equation of motion obtained
from the replacement $\Phi^{M'}_{\phi}\rightarrow
\Phi^{M'}_{\phi}+\Phi^{M'}_{\theta}$.
Expanding $\Phi^{M'}_{\theta}$ in powers of the spectral parameter $\lambda$, the infinite tower $\{\Phi^{M'}_{\theta_{n}}\}$ can be expressed recursively as
\begin{align}
\partial_u\Phi^{M'}_{\theta_{n}}=\partial_{\bar{w}}\Phi^{M'}_{\theta_{n-1}}+[\partial_u\Phi^{M'}_{\phi},\Phi^{M'}_{\theta_{n-1}}]^{M},
\qquad 
\partial_w\Phi^{M'}_{\theta_{n}}=\partial_{v}\Phi^{M'}_{\theta_{n-1}}+[\partial_w\Phi^{M'}_{\phi},\Phi^{M'}_{\theta_{n-1}}]^{M},
\label{HierarchyDefDefPleba}
\end{align}
with $\Phi^{M'}_{\theta_{-1}}=0$. Similar to the deformed self-dual Yang-Mills infinity tower (\ref{HierarchyDefSDYM}), the infinity tower (\ref{HierarchyDefDefPleba}) satisfies the compatibility condition $\partial_{w}\partial_{u}\Phi^{M'}_{\theta_{n}}=\partial_{u}\partial_{w}\Phi^{M'}_{\theta_{n}}$. However, also similarly to the deformed self-dual Yang-Mills theory, the integrability construction fails because the Jacobi identity of the Moyal bracket applied to deformed vector fields fails. \\

In this section, we have demonstrated the existence of a family of
generalised self-dual gauge and gravity theories obeying double copy
relationships. The key ingredients were the expression of such
theories as integral equations whose kernels manifestly contain
products of structure constants, and the introduction of new such
constants -- e.g., those of Moyal-deformed diffeomorphism algebra,
$X^M(p_1,p_2)$ -- that could be used as appropriate building blocks. In
all cases, we could show that the position space equations of motion
resulted from a Lax pair type of condition, although this was not always
sufficient to guarantee integrability. The Moyal deformation
considered above is in fact only one possible generalisation of the heavenly double copy between self-dual theories: we examine another in the following section.

\section{Generalising to the diffeomorphism algebra}
\label{sec:diff}

We will now, operating with a similar motive as in the previous section, consider a second -- independent -- generalisation of our usual self-dual theories. That is, we will consider a less restrictive gauge group, by relaxing the requirement that the kinematic group (SDiff) preserves the volume form, in this case the $(u,w)$-area form. This gives instead the group of full two-dimensional diffeomorphisms (Diff) in the $(u,w)$ plane. As we will see, this Diff group is in turn related to hyper-Hermitian manifolds, thus going beyond the hyper-K\"{a}hler structures in conventional self-dual gravity, discussed here in section~\ref{sec:Lax}.\\

Let us take a vector field $\Psi_{f}$ in the $(w,u)$ plane:
\begin{equation}
\Psi_{f}\equiv f^{A}\partial_{A}=f^{w}\partial_{w}+f^{u}\partial_{u},
\label{vectorPhi}
\end{equation}
where $A\in\{w,u\}$ and $f^{A}=f^{A}(u,v,w,\bar{w})$ depend on all the coordinates. In this form, $\Psi_{f}$ is an element of diff, the Lie algebra of Diff. The vector $\Psi_{f}$ satisfies the Jacobi identity
\begin{equation}
[[\Psi_{f},\Psi_{g}],\Psi_{h}]+[[\Psi_{g},\Psi_{h}],\Psi_{f}]+[[\Psi_{h},\Psi_{f}],\Psi_{g}]=0,
\label{JacobiDiff1}
\end{equation}
for the functions $f^{A},g^{A},h^{A}$. To make the connection with the previous sections, we need the kinematic object associated with the Lie algebra diff. To this end, let us define a vector in the direction $A$ and momentum $p$ by $\Psi_{p,A}\equiv e^{-ipx}\partial_{A}$. The commutator between two of such vectors is
\begin{equation}
[\Psi_{p,A},\Psi_{q,B}]=iY^{C}(q_{A},p_{B})\Psi_{p+q,C},
\end{equation}
where we define the kinematic object
\begin{eqnarray}
Y^{A}(p_{1C},p_{2B})\equiv p_{2B}\delta_{C}^{A}-p_{1C}\delta_{B}^{A}.
\end{eqnarray}
This satisfies $Y^{A}(p_{1C},p_{2B})=-Y^{A}(p_{2B},p_{1C})$ and also,
as we will see shortly, a Jacobi identity. As a consequence, it will
be interpreted as a structure constant of the two-dimensional
diffeomorphism algebra (diff). Following eq. (\ref{JacobiDiff1}), we have the Jacobi relation
\begin{equation}
Y^{D}(q_{A},p_{B})Y^{E}(k_{D},(p+q)_{C})+Y^{D}(k_{B},q_{C})Y^{E}(p_{D},(q+k)_{A})+Y^{D}(p_{C},k_{A})Y^{E}(q_{D},(p+k)_{B})=0,
\end{equation}
for any $E$. The kinematic factor $Y^{A}(p_{1C},p_{2B})$, which is the structure constant for diff, will be our focus in this section.

\subsection{Hyper-Hermitian manifold}

The notion of a hyper-Hermitian manifold is a generalisation of the hyper-K\"{a}hler case, where the volume preserving condition $\mathcal{L}_{L}\upsilon=\mathcal{L}_{M}\upsilon=0$ is relaxed on the gravitational Lax pair (\ref{Lax:Gravity}). Taking the Lax pair (\ref{Lax:Gravity}), we can rewrite the condition $[L,M]=0$ as
\begin{equation}
[U,W]=0,\quad [U,V]+[\bar{W},W]=0, \quad [\bar{W},V]=0.
\label{CondVect}
\end{equation}
Then, ${\mathbf V}=\{U,V,W,\bar{W}\}$ forms a null tetrad for a hyper-Hermitian metric  on a four-dimensional complex manifold ${\cal M}$. Following \cite{Grant_1999,doi:10.1063/1.522947,Dunajski:1998nj}, we
can define the vectors
\begin{align}
U&=\partial_{u}, &
W&=\partial_{w}, \nonumber \\
V&=\partial_{v}+\partial_{w}\Psi_{f}, &
\bar{W}&=\partial_{\bar{w}}+\partial_{u}\Psi_{f}.\label{Vectors2}
\end{align}
As in previous sections, for notational brevity, we denote $\partial_{w}\Psi_{f}\equiv [\partial_{w},\Psi_{f}]$ and $\partial_{u}\Psi_{f}\equiv [\partial_{u},\Psi_{f}]$.
The tetrad ${\mathbf V}$ satisfies eq.~(\ref{CondVect}) if and only if
$f^{A}$ satisfies
\begin{equation}
\partial^{2}f^{A}+\{f^{B},\partial_{B}f^{A}\} =0.
\label{HH}
\end{equation}
It was shown in ref.~\cite{Dunajski:1998nj} that this system describes
an hyper-Hermitian manifold. In momentum space, eq.~(\ref{HH})
implies the integral equation
\begin{equation}
f^{A}(k)=\frac{1}{2} \int \dd p_{1}\dd p_{2}\frac{\del (p_{1}+p_{2}-k)}{k^{2}}X (p_{1},p_{2})Y^{A}(p_{1C},p_{2B})f^{B}(p_{1})f^{C}(p_{2}),
\label{MomentumHH}
\end{equation}
thus justifying our above remark relating
the Diff group to hyper-Hermitian manifolds.\\

Before proceeding, let us point out that Hyper-Hermitian geometries can be more formally defined as
follows. Let $\mathcal{M}$ be a four-dimensional manifold and $g$ be a
Riemannian metric on $\mathcal{M}$. If $\mathcal{M}$ is equipped with
three complex structures $I,J,K$ satisfying the algebra of
quaternions, i.e., $IJ=-JI=K$, and $g$ is a Hermitian metric for the
three complex structures, then $\mathcal{M}$ is hyper-Hermitian. If,
besides this, the three K\"ahler forms are closed,
$d\Omega_{I}=d\Omega_{J}=d\Omega_{K}=0$, then $\mathcal{M}$ is
hyper-K\"ahler. A hyper-Hermitian manifold has a self-dual Weyl
tensor. On the other hand, a hyper-K\"ahler manifold has
a self-dual Weyl tensor and also a vanishing Ricci tensor.\footnote{These statements apply to our convention on (anti-)self-duality. Literature following different conventions may take the Weyl tensor for these manifolds to be anti-self-dual instead.}\\

Similarly to the arguments in section~\ref{sec:Moyal}, the fact that
the kernel of the integral equation of eq.~(\ref{MomentumHH}) involves
a product of structure constants immediately furnishes it with a
double copy interpretation. In particular, we note that
eq. (\ref{MomentumHH}) can be obtained via the double-copy-like
replacement
\begin{eqnarray}
f^{abc}\rightarrow Y^A(p_{1C},p_{2B}),
\label{DoubleCopyDiff}
\end{eqnarray}
in the structure constants of the colour Lie algebra in
eq. (\ref{YMsol}). Furthermore, eq. (\ref{MomentumHH}) then shows that
hyper-Hermitian manifolds are governed by a product of kinematic Lie
algebras, namely 
$$
\text{sdiff}\; \times\; \text{diff}.
$$

Now, the Lax pair $L$ and $M$ associated with the vectors (\ref{Vectors2}) for an hyper-Hermitian manifold
is given by
\begin{equation}
L=\partial_{u}-\lambda(\partial_{\bar{w}}+\partial_{u}\Psi_{f}),\qquad\qquad M=\partial_{w}-\lambda(\partial_{v}+\partial_{w}\Psi_{f}),
\label{LaxHH}
\end{equation}
and the eq. (\ref{HH}) arises from the condition $[L,M]=0$. We can recover conventional self-dual gravity if we impose the condition $\mathcal{L}_{L}\upsilon=\mathcal{L}_{M}\upsilon=0$ on the Lax pair (\ref{LaxHH}): there must then exist a function $\phi$ such that $f^{w}=-\partial_{u}\phi$ and $f^{u}=\partial_{w}\phi$, such that eq.~(\ref{HH}) reduces to the single Pleba\'nski equation (\ref{Plebanski}).\\

Similarly to the previous cases, we can construct a hierarchy of
linearised solutions. The linearised hyper-Hermitian equation, for
$\Psi_{f}\rightarrow\Psi_{f}+\Psi_{g}$, is
\begin{equation}
\partial^{2}\Psi_{g}+[\partial_{w}\Psi_{f},\partial_{u}\Psi_{g}]+[\partial_{w}\Psi_{g},\partial_{u}\Psi_{f}]=0.
\label{LinearisedHH}
\end{equation}
Expanding $\Psi_{g}$ in powers of $\lambda$, the infinite tower of
linearised solutions $\{\Psi_{g_{n}}\}$ to (\ref{LinearisedHH}) is
recursively expressed as
\begin{equation}
\partial_{u}\Psi_{g_{n}}=\partial_{\bar{w}}\Psi_{g_{n-1}}+[\partial_{u}\Psi_{f},\Psi_{g_{n-1}}],\qquad \partial_{w}\Psi_{g_{n}}=\partial_{v}\Psi_{g_{n-1}}+[\partial_{w}\Psi_{f},\Psi_{g_{n-1}}],
\end{equation}
with $\Psi_{g_{-1}}\equiv0$. This $\Psi_{g_{n}}$ satisfies the compatibility condition $\partial_{u}\partial_{w}\Psi_{g_{n}}=\partial_{w}\partial_{u}\Psi_{g_{n}}$, and the tower solves eq.~(\ref{LinearisedHH}). This relies on the Jacobi identity (\ref{JacobiDiff1}).\\

In addition to the structure constants of diff considered above, we could also introduce a Moyal deformation of sdiff.\footnote{Seen in a different way, we could start from eq.~(\ref{DSDYMMomentum}) and make the replacement (\ref{DoubleCopyDiff}).}
This results
in a deformed version of the hyper-Hermitian equation (\ref{HH}).  The
integral form of the equation of motion reads
\begin{equation}
f^{A}(k)=\frac{1}{2} \int dp_{1}dp_{2}\frac{\del (p_{1}+p_{2}-k)}{k^{2}}X^{M}(p_{1},p_{2})Y^{A}(p_{1C},p_{2B})f^{B}(p_{1})f^{C}(p_{2}),
\label{DHHMommentum}
\end{equation}
and it is straightforward to obtain the equation in position space,
\begin{equation}
\partial^{2}f^{A}+\{f^{B},\partial_{B}f^{A}\}^{M} =0.
\label{DHH}
\end{equation}
Using the Lax pair eq. (\ref{LaxHH}), this deformed equation arises from the condition $[L,M]^{M}=0$. The linearised deformed hyper-Hermitian equation, for
$\Psi_{f}\rightarrow\Psi_{f}+\Psi_{g}$, is
\begin{equation}
\partial^{2}\Psi_{g}+[\partial_{w}\Psi_{f},\partial_{u}\Psi_{g}]^{M}+[\partial_{w}\Psi_{g},\partial_{u}\Psi_{f}]^{M}=0.
\label{LinearisedDefHH}
\end{equation}
Expanding $\Psi_{g}$ in power of $\lambda$ and using the Lax equation, the infinite tower $\{\Psi_{g_{n}}\}$ is recursively expressed as
\begin{equation}
\partial_{u}\Psi_{g_{n}}=\partial_{\bar{w}}\Psi_{g_{n-1}}+[\partial_{u}\Psi_{f},\Psi_{g_{n-1}}]^{M},\qquad \partial_{w}\Psi_{g_{n}}=\partial_{v}\Psi_{g_{n-1}}+[\partial_{w}\Psi_{f},\Psi_{g_{n-1}}]^{M},
\label{HierarchyDefHH}
\end{equation}
with $\Psi_{g_{-1}}\equiv0$. In this naive hierarchy we are using the Moyal bracket on non-Hamiltonian vector fields, and similarly to the deformed self-dual Yang-Mills and doubly deformed Pleba\'nski cases, the Jacobi identity fails, and hence integrability is lost.\footnote{A
  different version of a deformed hyper-Hermitian structure is
  presented in \cite{doi:10.1063/1.2777008}.  That particular example
  formally retains integrability of the theory, although a twistor
  description is missing. Following \cite{Ovsienko1998DeformingTL},
  they make a deformation of the standard homomorphism between
  diff($S^{1}$) and the Poisson algebra on the cotangent bundle
  $T^{\ast}S^{1}$. It would be interesting to know whether these ideas fit into an expanded double copy web of theories.}\\

\subsection{Other Diff theories}

Motivated by the appearance of the Diff group, a natural question is:
what is the `gauge theory' with kinematic algebra diff? This gauge
theory is the single copy of the hyper-Hermitian
eq. (\ref{MomentumHH}) (or its deformed version (\ref{DHHMommentum})),
via the replacement $X(p_{1},p_{2})\rightarrow f^{abc}$. The
`Diff-gauge' theory is described by the equation
\begin{equation}
\Phi^{aA}(k)=\frac{1}{2}\int \dd p_{1}\dd p_{2}\frac{\del (p_{1}+p_{2}-k)}{k^{2}}f^{abc}Y^{A}(p_{1C},p_{2B})\Phi^{bB}(p_{1})\Phi^{cC}(p_{2}),
\label{GtimesDiffMomentum}
\end{equation} 
for $A=u,w$. It is straightforward to express it in position space as
\begin{equation}
\partial^{2}\Phi^{A}+i[\Phi^{B},\partial_{B}\Phi^{A}]=0.
\label{GtimesDiff}
\end{equation}
We can recover the SDYM eq. (\ref{YMEOM}) if we impose on $\Phi^A$ the condition
$\Phi^{w}=-\partial_{u}\Phi$ and $\Phi^{u}=\partial_{w}\Phi$. Of course, this condition is similar to imposing the volume
preserving condition in the hyper-Hermitian case. \\

Finally, we can make the replacement (\ref{DoubleCopyDiff}) in
eq.~(\ref{GtimesDiffMomentum}) and obtain a theory with two copies of
the diff algebra. In momentum space these equations are
\begin{equation}
f^{AB}(k)=\frac{1}{2}\int \dd p_{1}\dd p_{2}\frac{\del (p_{1}+p_{2}-k)}{k^{2}}Y^{A}(p_{1E},p_{2C})Y^{B}(p_{1F},p_{2D})f^{CD}(p_{1})f^{EF}(p_{2}),
\label{DiffTimesDiffMomentum}
\end{equation}
for $A,B,C,D=u,w$, whereas in position space, we have
\begin{equation}
\partial^{2}f^{AB}+f^{CD}(\partial_{C}\partial_{D}f^{AB})-(\partial_{C}f^{AD})(\partial_{D}f^{CB})=0.
\end{equation}

Equations~(\ref{GtimesDiffMomentum}) and~(\ref{DiffTimesDiffMomentum})
complete our web of theories, by combining all of the modifications of
self-dual Yang-Mills that we have considered. Again, the physical relevance of the new theories on their own -- outside the remit of the double copy -- is not clear, and it would be interesting to explore them further. 
\\


We pointed out early on the well-known fact that it is possible to recover self-dual
gravity (hyper-K\"ahler manifold) as a symmetry reduction of the
self-dual Yang-Mills theory, considering the gauge group as the
area-preserving diffeomorphisms group on the two-surface $\Sigma$,
SDiff($\Sigma$)
\cite{Plebanski:1995gk,plebanski1994selfdual,park19922d}, and that this
can be used to build hyper-K\"ahler metrics from solutions of
two-dimensional reduced SDYM equations
\cite{Chacon:2018nff,Dunajski_1998,plebanski1996principal}. In a
similar form, as seen in this section, we can obtain the
hyper-Hermitian equations taking the gauge group as the
area-preserving diffeomorphisms group in the Diff-gauge equations. It
would be interesting to see if it is possible to build hyper-Hermitian
metrics from solutions of two-dimensional reduced Diff-gauge
equations.\\

\section{Summary of results}
\label{sec:summary}


\begin{table}[ht]
\begin{tabular}{ |p{2cm}||p{1.5cm}|p{3.1cm}|p{3.9cm}|p{3.5cm}| }
 \hline
 $\times$ & $f^{abc}$ (G) & \cellcolor{Gray} $X(p_{1},p_{2})$ (sdiff) & $X^{M}(p_{1},p_{2})$ (Moyal) & $Y^{A}(p_{1B},p_{2C})$ (diff) \\ 
 \hline \hline
$f^{abc}$ (G) & Biadjoint scalar & \cellcolor{Gray} SDYM & Moyal-deformed SDYM & Diff-Gauge\\		 
\hline 
\cellcolor{Gray} $X(p_{1},p_{2})$ (sdiff) & \cellcolor{Gray} -- & \cellcolor{Gray} SDG Pleba\'nski (Hyper-K\"ahler)& \cellcolor{Gray} Moyal-deformed SDG (Hyper-K\"ahler) & \cellcolor{Gray} Hyper-Hermitian\\ 
\hline 

$X^{M}(p_{1},p_{2})$ (Moyal) & -- & \cellcolor{Gray} -- & Moyal$\times$Moyal SDG & Moyal-deformed Hyper-Hermitian \\ 
\hline 
$Y^{A}(p_{1B},p_{2C})$ (diff) & -- & \cellcolor{Gray} -- & -- & Diff$\times$Diff Gravity \\ 
\hline 
\end{tabular}
\caption{The web of theories studied in this paper. Each theory has an
  integral equation containing two structure constants, appearing on
  the left-hand and upper entries of the table. We represented in grey the theories that are integrable.
  }
\label{tab:Theories}
\end{table}

%

Throughout this paper, we have established a set of theories that
generalise self-dual Yang-Mills and gravity. The momentum space
equation for each theory contains a manifest product of structure
constants associated with colour and / or kinematic algebras, thus
allowing theories within the set to be related via the double copy. We summarise the complete set in table \ref{tab:Theories},
where the first column (consisting of choosing two potentially
distinct colour algebras) corresponds to biadjoint scalar theory. The
second column contains the the usual SDYM and SDGR theories (for a
hyper-K\"ahler manifold), whose double copy was discussed in detail in
ref.~\cite{Monteiro:2011pc}. The third column corresponds to our first
generalisation of the double copy procedure (section~\ref{sec:Moyal}),
in which one of the structure constants is taken to be a Moyal deformation of the kinematic
structure constant. This leads to single Moyal
deformations of SDYM and SDGR, and also a doubly deformed gravity
theory, whose equation of motion contains a product of two Moyal
kinematic factors. We should point out the (non-integrable) Moyal-deformed SDYM case considered in the table, which is the one that straightforwardly matches the double copy structure, is not the same as the more conventional Moyal deformation of SDYM commonly considered in the literature, as discussed in section~\ref{sec:YMdeform}.
The second generalisation considers the kinematic
algebra diff, where the area preserving condition has been relaxed. A
product with sdiff results in a Hyper-Hermitian manifold. In section
\ref{sec:diff}, we have studied this as well as its Moyal deformation,
a Diff-gauge theory, and a double-Diff gravity, as recapped in the
fourth column of Table~\ref{tab:Theories}.\\

The entries of Table~\ref{tab:Theories} identified in grey correspond to integrable theories. All such theories have $X(p_1,p_2)$ as one of the double copy factors, and this factor is at the origin of their integrability. The equation of motion in momentum space reads 
\begin{equation}
\Upsilon_{\rho}(p)=\frac{1}{2}\int \dd p_{1}\dd p_{2}\frac{\del(p_{1}+p_{2}-p_{})}{p^{2}}\,X(p_{1},p_{2})\;\Pi_{{1}\,{2}}\;\Upsilon_{\rho_1}(p_{1})\Upsilon_{\rho_2}(p_{2}),
\label{GeneralEMMomentum}
\end{equation}
where $\Pi_{{1}\,{2}}$ provides the other structure constant. In coordinate space, we have
\begin{equation}
\partial^{2}\Upsilon_{\rho}+[\partial_{w}\Upsilon_{\rho},\partial_{u}\Upsilon_{\rho}]=0.
\label{GeneralEPosition}
\end{equation}
Pleasingly, all the integrable theories can be written in a similar manner,
where $\Upsilon_{\rho}$ is $\Phi=\Phi^aT^a$ for self-dual Yang-Mills, is $\Phi_\phi$ for self-dual gravity, is  $\Phi_\phi^M$ for singly Moyal-deformed self-dual gravity, and is $\Psi_f$ for hyper-Hermitian gravity. Their Lax pairs can be expressed as 
\begin{equation}
L=\partial_{u}-\lambda(\partial_{\bar{w}}+\partial_{u}\Upsilon_{\rho}),\qquad M=\partial_{w}-\lambda(\partial_{v}+\partial_{w}\Upsilon_{\rho}),
\label{GeneralLax}
\end{equation}
and the equation of motion follows from the condition $[L,M]=0$. Integrability follows from the existence of an infinite hierarchy of conserved currents, associated to an infinite hierarchy of solutions to the linearised equation of motion. The latter is  obtained by making the replacement
$\Upsilon_{\rho}\rightarrow\Upsilon_{\rho}+\Upsilon_{\alpha}$, in eq.
(\ref{GeneralEPosition}):
\begin{equation}
\partial^{2}\Upsilon_{\alpha}+[\partial_{w}\Upsilon_{\rho},\partial_{u}\Upsilon_{\alpha}]^{}+[\partial_{w}\Upsilon_{\alpha},\partial_{u}\Upsilon_{\rho}]^{}=0.
\label{GeneralLinear}
\end{equation}
The hierarchy of solutions $\{\Upsilon_{\alpha_{n}}\}$ arises via
\begin{equation}
\partial_{u}\Upsilon_{\alpha_{n}}=\partial_{\bar{w}}\Upsilon_{\alpha_{n-1}}+[\partial_{u}\Upsilon_{\rho},\Upsilon_{\alpha_{n-1}}]^{},\qquad \partial_{w}\Upsilon_{\alpha_{n}}=\partial_{v}\Upsilon_{\alpha_{n-1}}+[\partial_{w}\Upsilon_{\rho},\Upsilon_{\alpha_{n-1}}]^{},
\end{equation}
where $\Upsilon_{\alpha_{-1}}\equiv0$. In this hierarchy, $\Upsilon_{\alpha_{n}}$ is guaranteed to satisfy the linearised equation of motion if $\Upsilon_{\alpha_{n-1}}$ does. \\

Several non-integrable theories in Table~\ref{tab:Theories} naively share some of these features, at least those which possess $X^M(p_1,p_2)$ as one of the double copy factors (we exclude here the case where $X(p_1,p_2)$ is the other factor, where the theory is integrable). Then eq.~\eqref{GeneralEPosition} changes only by employing a Moyal-deformed bracket, that is, we make the substitution $[\cdot,\cdot]\,\to\,[\cdot,\cdot]^M$. The following equations also appear to tell an analogous story with the same substitution. However, crucially, the Moyal-deformed bracket fails to obey the Jacobi identity, and the steps leading to the conclusion of integrability fail.  \\

\section{Conclusions}
\label{sec:conclude}

In this paper, we have constructed new examples of gauge and gravity
theories satisfying manifest double copy relations, motivated by
previous work in the self-dual sector of Yang-Mills theory and
gravity~\cite{Monteiro:2011pc}. Each of our theories has an integral
momentum-space equation containing a product of two (potentially different) sets of
structure constants, associated with colour and / or kinematic degrees
of freedom. We have considered two generalisations of
ref.~\cite{Monteiro:2011pc}, namely the use of Moyal deformations of
the kinematic sector, and also the use of a full two-dimensional
diffeomorphism group rather than its more restricted area-preserving
counterpart. This gives new kinematic structure constants, and our web
of theories -- summarised in table~\ref{tab:Theories} -- consist of
all possible combinations of these various building blocks.\\

We have also studied the integrability properties of the web of theories. This aspect was not considered in ref.~\cite{Monteiro:2011pc}, and we discussed how the well-known integrability constructions (e.g., Lax pairs) in self-dual Yang-Mills theory and self-dual gravity are related in a manner expected by the double copy. More generally, for the web of theories summarised in table~\ref{tab:Theories}, the only integrable theories are those for which at least one of the two sets of structure constants is $X(p_1,p_2)$, associated to sdiff, i.e., the Lie algebra of area preserving diffeomorphisms. These include self-dual Yang-Mills theory, self-dual gravity and its single Moyal deformation, and hyper-Hermitian gravity (an extension of self-dual gravity associated to hyper-Hermitian manifolds instead of hyper-K\"ahler manifolds). In those cases, it is possible to construct a Lax pair, and to relate it to an infinite hierarchy of linearised symmetries, expressed as conserved currents. When none of the two sets of structure constants was associated to sdiff, the integrability construction failed. In these cases, we may still have a presumptive Lax pair interpretation, and a bracket based on a deformation of the
standard Lie or Poisson brackets, but this bracket  fails to obey the Jacobi relation. This agrees with previous literature, e.g., ref.~\cite{Strachan:1996gx} regarding Moyal-deformed brackets. We have the interesting
consequence that singly Moyal-deformed self-dual gravity can be
thought of as ``inheriting'' its integrability from one of its
constituent gauge theories under the double copy, i.e., from the one with kinematic algebra $X(p_1,p_2)$ and not the one with kinematic algebra $X^M(p_1,p_2)$. The picture that emerges is that, when integrability exists, it does so because at least one of the sides of the double copy is integrable. In the cases at hand, all integrable theories are related to self-dual Yang-Mills theory by substitution of $f^{abc}$ by other structure constants, which can be thought of as arising from a `symmetry reduction'. Hence, the class of integrable theories considered here is consistent with Ward's conjecture \cite{Ward:1985gz,Mason:1991rf}.  \\

There are a number of possible avenues for further work. Firstly, it would useful to establish the possible physical interpretation and applications of the theories that have been introduced in order to complete the web obtained via the double copy. This includes the question of whether or not they can be furnished with a geometric interpretation. Finally, our hope is that the results of our study may stimulate the construction of yet more double copy examples, which may possibly in turn enhance our understanding of conventional Yang-Mills and gravity beyond the self-dual sector.


\section*{Acknowledgements}

We are grateful to Maciej Dunajski and Maciej Przanowski for useful discussions. This work has been supported by the UK Science and Technology Facilities Council (STFC) Consolidated Grant ST/P000754/1 ``String theory, gauge theory and duality'', and by the European Union Horizon 2020 research and innovation programme under the Marie Sk\l{}odowska-Curie grant agreement No. 764850 ``SAGEX''. EC is supported by the National Council of Science and Technology (CONACYT). AL is supported by the Department of Energy under Award Number DE-SC0009937. RM is supported by a Royal Society University Research Fellowship.

\bibliography{refs}
\end{document}